\documentclass{article}
\usepackage{epsf}
\usepackage{psfig}
\usepackage{psfrag}
\usepackage{epsfig}
\usepackage{float}
\newcommand{\beq}{\begin{equation}}
\newcommand{\eeq}{\end{equation}}
\newcommand{\beqa}{\begin{eqnarray}}
\newcommand{\eeqa}{\end{eqnarray}}

\newcommand{\vs}{\vspace{-0.15cm}}

\begin{document}

\thispagestyle{empty}

\par
\topmargin=-1cm      

{ \small

\hfill \noindent{\tiny FZJ-IKP(TH)-2002-12} 

}

\thispagestyle{empty}

\vspace{80.0pt}

\begin{centering}
{\Large\bf Near threshold neutral pion electroproduction on deuterium\\[0.3em] 
in chiral perturbation theory}\footnote{Work
supported in part by Deutsche Forschungsgemeinschaft under contract
no. Me864-16/2.}\\

\vspace{30.0pt}
{\bf H.~Krebs}$^1$,
{\bf V.~Bernard}$^2$,
{\bf Ulf-G.~Mei{\ss}ner}$^{1,3}$\\
\vspace{20.0pt}

{\sl $^{1}$Institut f\"ur Kernphysik (Theorie), Forschungszentrum J\"ulich}\\
{\sl D-52425 J\"ulich,
Germany} \\
{\it E-mail addresses: h.krebs@fz-juelich.de, u.meissner@fz-juelich.de}\\

\vspace{15.0pt}

{\sl $^{2}$Laboratoire de Physique Th\'eorique,
Universit\'e Louis Pasteur} \\
{\sl  F-67037 Strasbourg Cedex 2, France} \\
{\it E-mail address: bernard@lpt6.u-strasbg.fr}\\

\vspace{15.0pt}

{\sl $^{3}$Institut f\"ur Theoretische Physik,
  Karl-Franzens-Universit\"at Graz}\\
{\sl A-8010 Graz, Austria} \\

\end{centering}
\vspace{20.0pt}
\begin{center}
\begin{abstract}
\noindent
Near threshold neutral pion electroproduction on the deuteron is studied in
the framework of baryon chiral perturbation theory at next--to--leading 
order in the chiral expansion.  We develop the multipole decomposition
for pion production off spin-1 particles appropriate for the threshold
region. The existing data at photon virtuality $k^2 = -0.1\,$GeV$^2$
can be described satisfactorily. Furthermore, the prediction for the 
S--wave multipole $E_d$ at the photon point is in good agreement with
the data. 
\end{abstract}

\vspace*{50pt}
{\small

}
\vfill
\end{center}

\newpage

\section{Introduction}
\def\theequation{\arabic{section}.\arabic{equation}}
\setcounter{equation}{0}
\label{sec:intro}

Pion photo-- and electroproduction off single nucleons in the
threshold region can be considered one of the best testing grounds for
our understanding of the chiral pion--nucleon dynamics resulting from
the symmetry structure of QCD (for a recent status report, see
e.g. \cite{ulf}). In the absence of neutron targets, it is mandatory
to consider pion production off light nuclei which also leads
to the consideration of interesting aspects related to 
few--nucleon dynamics. In this paper, we consider pion
electroproduction on the deuteron above
threshold extending our previous work \cite{BKM}.
This is mandated by the following developments: First, the threshold
results obtained in \cite{BKM} can not be directly compared to the
data. Furthermore, the important single scattering contribution was
not calculated to fourth order (which is mandatory to describe
the elementary process with sufficient accuracy)  but simply shifted
to its value at the photon point. Such a procedure is only well 
controlled for the transverse multipole. Second, coherent neutral
pion production off deuterium at photon virtuality $k^2 = -0.1\,$GeV$^2$ has been
measured and  analyzed at MAMI \cite{Ewald},
and these data show a significant discrepancy in the dominant
longitudinal cross section (S--wave multipole) from the
prediction of \cite{BKM}. In addition, new measurements of
pion electroproduction off the proton at MAMI \cite{Merkel}
at the lower photon virtuality of $k^2 = -0.05\,$GeV$^2$ have 
led to intriguing results that
can neither by explained in chiral perturbation theory nor
with any sophisticated model (note the very unusual values for
certain P--waves given in that paper). Here, we want to improve the
calculation for coherent pion production off the deuteron in two ways. First,
in the single scattering contribution we include the full fourth
order result for the transverse and longitudinal S--waves,
with its parameters fixed from recent data on neutral pion
photoproduction and the older  NIKHEF \cite{benno} and MAMI
\cite{distler} measurements for
electroproduction off the proton at $k^2 = -0.1\,$GeV$^2$.
Note that it was recently shown that in the case of neutral
pion photoproduction the fourth order corrections to the 
P--wave multipoles are fairly small \cite{BKMa}. A similar 
analysis for electroproduction is not yet available, it is,
however, conceivable that similar trends will persist in that case.
Second, we calculate above threshold, which leads to a considerable
complication in terms of the multipole expansion. While this formalism
has already been developed in \cite{Aren1,Aren2}, we present here a
new form particularly suited for the threshold region and that most
closely resembles the single nucleon multipole expansion. We restrict 
ourselves to S-- and P--waves and evaluate the three--body corrections
(or meson--exchange currents) to third order in the chiral expansion.
We include, however, the pion mass difference which is formally of
higher order but constitutes the dominant isospin breaking effect.
The single scattering part for the proton has been fixed before, 
and can be reliably estimated for the neutron using resonance saturation
at the photon point.
However, at finite photon virtuality the situation is less clear and
we do not want to rely on the resonance saturation hypothesis. 
We therefore perform two types of fits. In the minimal fit we employ
resonance saturation for a dimension four LEC (there are in principle
two LECs but their sum is constrained by a low--energy theorem \cite{BKMe})
and use  the longitudinal deuteron multipole $L_d$ as extracted from the MAMI data 
to pin down one parameter related to a particular dimension five 
operator \cite{BKMe}. 
In a second scenario, we do not use resonance saturation
and thus have two free parameters related to the polynomial part of
$L_{0+}^n$ which we determine
from a best fit to the measured total cross sections at low pion excess energies.
This still leaves sufficient predictive power since we can compare directly
with the measured differential cross sections or the extracted S--wave
cross section $a_{0d}$. For doing
that, one has to select a deuteron wave function. In \cite{BKM}, we
employed the hybrid approach using various high precision wave
functions together with the chirally expanded interaction kernel.
Here, we also improve on that aspect using recently obtained 
precise effective field theory wave functions that are consistent 
with the power counting of the kernel 
\cite{EGMII,EGME,Eetal}.\footnote{We have also performed calculations
using  precise phenomenological wave functions as a check. None of
the results shown later depend on the choice of wave function.}
We will demonstrate that this improved calculation is in fair agreement with
the MAMI deuteron data at $k^2 = -0.1\,$GeV$^2$\cite{Ewald}, 
thus solving one apparent discrepancy and deepening the mystery
surrounding the data of Ref.\cite{Merkel}. Needless to say that
a separate investigation of this second puzzle is urgently called 
for but should not be a topic of the present paper.

\medskip\noindent
This paper is organized as follows. In Section~\ref{sec:mult} we
discuss the multipole decomposition for neutral pion electroproduction
off a spin--1 target. In Section~\ref{sec:EFT}
we briefly review the effective Lagrangian underlying the calculation and the
standard power counting formulas.  In Section~\ref{sec:anat}
the calculation of the various contributions to the transition current
(single scattering and three--body terms) 
is outlined. Section~\ref{sec:res} contains the results and discussions thereof.  
A brief summary and outlook is given in Section~\ref{sec:summ}.
The appendices include our conventions and give many more details 
on the calculations.

\vfill
\section{Multipole decomposition}
\def\theequation{\arabic{section}.\arabic{equation}}
\setcounter{equation}{0}
\label{sec:mult}

The main part of this section is concerned with the multipole
decomposition for the process $\gamma^* d \to \pi^0 d$. To 
develop this, we heavily rely on the work of Arenh\"ovel 
\cite{Aren1,Aren2} for the classification of the operator
basis, construction of invariant amplitudes and the calculation of
observables. However, in his work the main emphasis was put on the 
helicity basis. A formal proof of the equivalence between
the multipole expansion used here and the one of Arenh\"ovel
is given in appendix~\ref{app:equiv}. We also summarize some basic formulae 
to calculate observables from the multipoles.

\medskip\noindent
The invariant matrix element for the process $\gamma^\star (k) + d (p_d)\to
\pi^0 (q) + d (p_d')$, where $\gamma^*$ denotes the virtual photon with
virtuality $k^2 \le 0$, $d$ the deuteron and $\pi^0$ the neutral
pion with four--momentum $q_\mu = (\omega, \vec{q}\,)$, 
can be expressed in terms of 13 invariant functions,
\beq 
{\cal M}^\lambda  = \sum_{i=1}^{13} {\cal O}_i^\lambda  \, F_i~,
\eeq
where the 13 operators ${\cal O}_i^\lambda$ 
are expressed in terms of combinations of the
direction of the photon three--momentum $\hat k$, the photon 
polarization vector $\vec{\varepsilon}\,^\lambda$, the
direction of the pion three--momentum $\hat q$ and the
deuteron spin vector $\vec S$. Here, $\lambda$
denotes the helicity of the in--coming photon, with $\lambda = 0, \pm 1$.
This non--relativistic form is most appropriate for near threshold 
production. The explicit form of the ${\cal O}_i^\lambda$, 
first written down in~\cite{Aren1}, is 
\beqa 
{\cal O}_1^\lambda 
           &=& \vec{\varepsilon\,}^\lambda 
                \cdot (\hat k \times \hat q ) ~, \quad
{\cal O}_2^\lambda   = \vec{\varepsilon\,}^\lambda 
                \cdot (\hat k \times \hat q ) \,
              \vec{S} \cdot (\hat k \times \hat q ) ~,\quad
{\cal O}_3^\lambda   = \vec{\varepsilon\,}^\lambda  
               \cdot (\hat k \times (\hat k \times
\vec{S}))~, \nonumber \\
{\cal O}_4^\lambda  &=& \vec{\varepsilon\,}^\lambda 
                \cdot (\hat k \times (\hat q \times
\vec{S}))~,\quad
{\cal O}_5^\lambda   = \vec{\varepsilon\,}^\lambda 
                             \cdot (\hat k \times \hat q )
\, \hat{k}^{[2]} \cdot S^{[2]}~,\quad
{\cal O}_6^\lambda   = \vec{\varepsilon\,}^\lambda 
                            \cdot (\hat k \times \hat q )
\, [\hat{k} \times \hat q]^{[2]} \cdot S^{[2]}~,\nonumber \\
{\cal O}_7^\lambda   &=& \vec{\varepsilon\,}^\lambda 
                            \cdot (\hat k \times \hat q )
\,\hat{q}^{[2]} \cdot S^{[2]}~,\quad
{\cal O}_8^\lambda 
           = \vec{\varepsilon\,}^\lambda 
                    \cdot \left(\hat k \times [\hat k \times
{S}^{[2]}]^{[1]}\right)~, \quad
{\cal O}_9^\lambda 
           = \vec{\varepsilon\,}^\lambda 
                        \cdot \left(\hat k \times [\hat q \times
{S}^{[2]}]^{[1]}\right)~, \nonumber \\
{\cal O}_{10}^\lambda 
          &=& \vec{\varepsilon\,}^\lambda 
                           \cdot \hat k \, \hat k \cdot \vec S~,
\quad 
{\cal O}_{11}^\lambda 
           = \vec{\varepsilon\,}^\lambda 
                     \cdot \hat k \, \hat q \cdot \vec S~,
\quad 
{\cal O}_{12}^\lambda   = \vec{\varepsilon\,}^\lambda  
                                  \cdot \hat k \, \left[ ( \hat k
\times \hat q) \times \hat k\right]^{[2]} \cdot  S^{[2]}~, \\
{\cal O}_{13}^\lambda   &=& \vec{\varepsilon\,}^\lambda 
                       \cdot \hat k \, \left[ ( \hat k
\times \hat q) \times \hat q\right]^{[2]} \cdot  S^{[2]}~,\nonumber
\eeqa
with 
\beq
\left[ \vec{u} \times S^{[2]} \,\right]_{k}^{[1]} =
u_l^{\phantom{{[2]}}} \!\! S_{lk}^{[2]}~,
\eeq
and 
\beq
\left[ \vec{a} \times \vec{b} \,\right]_{ij}^{[2]} 
= \frac{1}{2} \left( a_i b_j + a_j b_i \right) - \frac{1}{3}
\delta_{ij} \vec{a} \cdot \vec{b} 
\eeq
the symmetric traceless tensor of second rank in standard
notation. The first nine of
these operators are transverse, whereas the other four are longitudinal.
Note also that we work in the Coulomb gauge $\varepsilon_0 =0$.
Furthermore, the $F_i$ are functions of three kinematical variables, 
we chose here to work with the pion energy $\omega$, the photon 
virtuality and  the scattering angle, $F_i = F_i (\omega , k^2, z)$.
More precisely, $\theta$ is the scattering angle in the $\pi^0 d$ 
center-of-mass system with $z = \cos \theta$. In what follows, we will
however not display these arguments. For later use, we define the
vector $\tilde{F}$ via
\beq
\tilde{F} = \left( F_1, F_2, F_3 , \ldots , F_{13} \right)~.
\eeq
Any given tree or loop graph can now be
expanded in this basis, and all observables can be expressed as
functions of the $F_i$. Explicit expressions can be
found in \cite{Aren2}.

\begin{figure}[htb]
   \vspace{0.5cm}\hspace{1cm}
   \epsfysize=4.5cm
   \epsffile{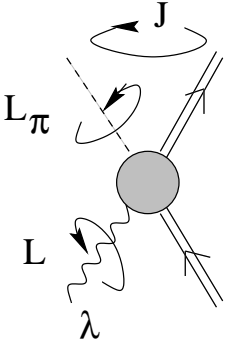}

\vspace{-2.9cm}

\hspace{6cm}
\parbox{9cm}{\caption{\label{fig:mult}
Graphical representation of the angular momenta involved in pion
electroproduction. The in-coming photon (wiggly line) 
has helicity $\lambda$ and
multipolarity $L$. The final $\pi^0 d$ state is characterized by the
total angular momentum $J$ and the pion (dashed line) angular momentum $L_\pi$.
The double line denotes the deuteron.}}
\end{figure}

\medskip\noindent
However, for the analysis of the data and the direct comparison
with theoretical predictions, it is advantageous to use a
multipole decompositon similar to the standard case of pion
production off a single nucleon. The pertinent method to do that
has been outlined a long time ago in~\cite{PK}, and we use that
formalism here to develop the multipole decomposition for our case.
As shown in Fig.~\ref{fig:mult}, a photon with helicity $\lambda$
and multipolarity $L$ produces the neutral pion. In the final--state
$\pi^0 d$ system, the pion has relative orbital angular momentum $L_\pi$,
which couples with the deuteron spin to the total angular momentum $J$.
\noindent
In the $\pi^0 d$ frame, the invariant matrix element for the unpolarized
case considered here can be written as
\beqa
{\cal M}^\lambda &=&  \sum_{m,m'} |m'\rangle \langle \,
\hat{q} \, m' \,| a_\mu^\lambda J^\mu_d (\omega,k^2) 
|\,\hat{k}\,\lambda\,m \rangle
\langle \lambda \, m\,|~, \nonumber \\
&=& \sum_{m,m'} |m'\rangle \, t_{m',\lambda,m} (\theta)\, \langle 
\lambda \, m\,|~,
\eeqa
where $a_\mu^\lambda = \varepsilon_\mu^\lambda -
(\varepsilon^\lambda_0 / k_0) \,  k_\mu$ 
is the transverse polarisation vector
and $J^\mu_d$ is the photon vector current
impinging on the deuteron and we have summed over the initial and
final--state deuteron magnetic quantum numbers. 
This matrix element can be expressed in terms
of electric $E$, magnetic $M$ and longitudinal $L$ multipoles 
as\footnote{Note the symbol $L$ is used for the multipolarity
and the longitudinal multipoles. However, no confusion can arise since
it is always obvious from the context what is meant.}
(note that we do not give the explicit dependence on the azimuthal
angle $\varphi$ here since we will only consider the transverse and the
longitudinal cross section in what follows, see also the discussion below)
\beq\label{MEl}
{\cal M}^\lambda = \sum_{m,m'} \sum_{L,L_\pi,J} \,
\hat{\theta}_{m',m}^{L_\pi,\lambda} \, D_{\lambda,m',m}^{L,L_\pi,J}
\, {\cal O}^{L,\lambda}_{L_\pi,J}~,
\eeq
with 
\beqa
\hat{\theta}_{m',m}^{L_\pi,\lambda} & = & |m'\rangle \langle m| \,
Y_{L_\pi, \lambda +m-m'} (z, \varphi = 0)~, \nonumber \\
D_{\lambda,m',m}^{L,L_\pi,J} &=& \langle L_\pi \, \lambda+m-m' \,1 \,
m' \,| \, J \, \lambda+ m \, \rangle 
\langle 1 \, m \, L \, \lambda | J \, \lambda+m \rangle~,
\nonumber \\
{\cal O}^{L,\lambda}_{L_\pi,J} &=& {4i \sqrt{2\pi} \hat{L} \over
  \hat{J}} \left[ \left( E^L_{L_\pi,J} + \lambda M^L_{L_\pi,J} \right)
\delta_{|\lambda |,1} + {\sqrt{-k^2} \over -k_0} \, L^L_{L_\pi,J}
\delta_{\lambda,0} \right]~.
\eeqa
The first factor is proportional to the angular momentum eigenfunction
in the final state (i.e. the appropriate spherical harmonics), the 
second term collects the pertinent Clebsch--Gordan coupling coefficients
and the third term contains the dynamical information in terms of
the multipoles. These multipoles depend on the pion energy and the
photon virtuality. They are characterized by three labels. The superscript
$L$ refers to the multipolarity while the lower indices $L_\pi$ and $J$
denote the orbital angular momentum and the total angular momentum of the
final pion--deuteron system. We furthermore use $\hat A= \sqrt{2A+1}$
for angular momentum eigenvalues. Note also that for real photons with
$k^2 = 0$ (photoproduction), the photon has no longitudinal components 
and thus there is no coupling to the longitudinal multipoles. One
can invert Eq.~(\ref{MEl}) and project out the multipoles, this gives
\beqa
E^L_{L_\pi,J} &=& \frac{1}{2} \left(1 -(-)^{L_\pi+L}\right)
{\sqrt{2\pi} \over 4i} {\hat L \over \hat J}  \sum_{m,m'}
D^{L,L_\pi,J}_{1,m',m} \, \int_{-1}^1 dz \, \langle m' | {\cal
  M}_1|m\rangle \, Y_{L_\pi,1+m-m'} (z, \varphi=0)~, \nonumber\\
M^L_{L_\pi,J} &=& \frac{1}{2} \left(1 +(-)^{L_\pi+L}\right)
{\sqrt{2\pi} \over 4i} {\hat L \over \hat J}  \sum_{m,m'}
D^{L,L_\pi,J}_{1,m',m} \, \int_{-1}^1 dz \, \langle m' | {\cal
  M}_1|m\rangle \, Y_{L_\pi,1+m-m'} (z, \varphi=0)~, \nonumber\\
L^L_{L_\pi,J} &=& \frac{1}{2} \left(1 -(-)^{L_\pi+L}\right)
{-k_0 \over \sqrt{-k^2}}
{\sqrt{2\pi} \over 4i} {\hat L \over \hat J}  \sum_{m,m'}
D^{L,L_\pi,J}_{0,m',m} \, \int_{-1}^1 dz \, \langle m' | {\cal
  M}_0|m\rangle \, Y_{L_\pi,m-m'} (z, \varphi=0)~, 
\eeqa
where due to parity, the sum $L+L_\pi$ has to be odd for the electric
and the longitudinal multipoles and even for the magnetic ones.
Since now for a given orbital angular momentum $L_\pi$ we have
the conditions $|L_\pi -1| \le J \le L_\pi+1$ and $|J-1| \leq L \leq
J+1$, parity allows for four different electric, four longitudinal
and five magnetic multipoles, which we collect in the nine--component 
transverse vector $\tilde{T}_{L_\pi}$,  
\beq
\tilde{T}_{L_\pi} = \left(   E_{L_\pi,L_\pi-1}^{L_\pi-1}, 
E_{L_\pi,L_\pi}^{L_\pi-1},   E_{L_\pi,L_\pi}^{L_\pi+1},
E_{L_\pi,L_\pi+1}^{L_\pi+1}, M_{L_\pi,L_\pi-1}^{L_\pi-2},
M_{L_\pi,L_\pi-1}^{L_\pi},   M_{L_\pi,L_\pi}^{L_\pi},
M_{L_\pi,L_\pi+1}^{L_\pi},   M_{L_\pi,L_\pi+1}^{L_\pi+2}
\right)~,
\eeq
and the four--component longitudinal vector $\tilde{L}_{L_\pi}$,
\beq
\tilde{L}_{L_\pi} = \left(   L_{L_\pi,L_\pi-1}^{L_\pi-1}, 
L_{L_\pi,L_\pi}^{L_\pi-1},   L_{L_\pi,L_\pi}^{L_\pi+1},
L_{L_\pi,L_\pi+1}^{L_\pi+1}
\right)~.
\eeq
These together define the multipole vector $\tilde{{\cal M}}_{L_\pi}$,  
\beq
\tilde{{\cal M}}_{L_\pi} = \left( \tilde{T}_{L_\pi}, \tilde{L}_{L_\pi}
\right)~,
\eeq
which has 13 components. It is straightforward albeit somewhat tedious
to work out the transformation matrices between the multipole basis
and the one spanned by the invariant functions $F_i$.
The multipoles can be obtained from the $F_i$ by a 13$\times$13
block--diagonal matrix, such that
\beq\label{matDE}
\tilde{{\cal M}}_{L_\pi}  = \int_{-1}^{+1} dz\,
\left( \begin{array}{cc} D_{L_\pi} (z) & 0 \\
               0 & E_{L_\pi} (z) \end{array} \right)~
 \tilde{F}~,
\eeq
where $D_{L_\pi}$ is a $9\times 9$ and  $ E_{L_\pi}$  a $4\times 4$ matrix.
The explicit representation of these matrices in terms of Legendre polynomials
is given in appendix~\ref{app:mult}. Similarly, the inverse transformation
is given in terms of a $9\times 9$ matrix, called $ G_{L_\pi}$ and a
$4\times 4$ matrix, denoted $ H_{L_\pi}$, as
\beq\label{matGH}
\tilde{F} = \sum_{L_\pi =0}^{\infty}
\left( \begin{array}{cc} G_{L_\pi} (z) & 0 \\
               0 & H_{L_\pi} (z) \end{array} \right)~
 \tilde{{\cal M}}_{L_\pi}~.
\eeq
The explicit form of these matrices is also given in  appendix~\ref{app:mult}.
As a non--trivial check we have shown that the
product of the two $13\times 13$ matrices in
Eqs.~(\ref{matDE},\ref{matGH}) is indeed the unit matrix.
Note that for the electric and the magnetic  multipoles  $L$ has to
be larger or equal to one (since $|\lambda | = 1$) and that $L_\pi =0$
or $L = 0$ implies $J = 1$. This reduces the number of allowed 
multipoles for $L_\pi \le 2$ and leads to the lowest permissible
value of $L_\pi$ for the various multipoles given in Table~\ref{tab:val}.
\renewcommand{\arraystretch}{1.4}
\begin{table}[htb]
\begin{center}
\begin{tabular}{|l||c|c|c|c|c|}
\hline
electric multipole  &  $E_{L_\pi,L_\pi-1}^{L_\pi-1}$ &
$E_{L_\pi,L_\pi}^{L_\pi-1}$ &  $E_{L_\pi,L_\pi}^{L_\pi+1}$ &
$E_{L_\pi,L_\pi+1}^{L_\pi+1}$ &  \\
\hline
lowest value of $L_\pi$ & 2 & 2 & 1 & 0 &  \\
\hline\hline
magnetic multipole  &  $M_{L_\pi,L_\pi-1}^{L_\pi-2}$ &
$M_{L_\pi,L_\pi-1}^{L_\pi}$ &  $M_{L_\pi,L_\pi}^{L_\pi}$ &
$M_{L_\pi,L_\pi+1}^{L_\pi}$ &  $M_{L\pi,L_\pi+1}^{L_\pi+2}$ \\
\hline
lowest value of $L_\pi$ & 3 & 1 & 1 & 1 & 0 \\
\hline\hline
longitudinal multipole  &  $L_{L_\pi,L_\pi-1}^{L_\pi-1}$ &
$L_{L_\pi,L_\pi}^{L_\pi-1}$ &  $L_{L_\pi,L_\pi}^{L_\pi+1}$ &
$L_{L_\pi,L_\pi+1}^{L_\pi+1}$ & \\
\hline
lowest value of $L_\pi$ & 2 & 1 & 1 & 0 &\\
\hline\end{tabular}
\caption{Lowest permissible value of $L_\pi$ for the various
  multipoles.
\label{tab:val}}
\end{center}
\end{table}

\medskip\noindent
The unpolarized differential cross section for neutral pion
electroproduction from a spin--1 target decomposes into four
structure functions. However, so far no data are available for
the small transverse--longitudinal and transverse--transverse
interference structure functions and we thus will also not
consider any azimuthal dependence here (i.e. angular dependence
between the scattering and the production plane). Then, the
differential cross section decomposes into a transverse and a
longitudinal part, the latter being multiplied by $\varepsilon_L$,
with $\varepsilon_L$ the longitudinal degree of virtual 
photon polarization which is related to the transverse one by
\beq
\varepsilon_L = -\frac{k^2}{k_0^2} \, \varepsilon ~,
\eeq
with the photon energy and momentum taken in the photon--deuteron
center-of-mass system. The explicit form of the
transverse cross section reads
\beqa
\sigma_T &=& {|\vec{q \,}| \over |\vec{k \,}|} \, 
\frac{1}{2} \sum_{\lambda = \pm 1} \, \frac{1}{3} \sum_{m',m}
\, \int d\Omega\, \left| t_{m',\lambda,m} (\theta) \right|^2 \nonumber \\
&=& 4\pi \,{|\vec{q \,}| \over |\vec{k \,}|} \, \frac{8}{3} \, 
\sum_{L_\pi,L,J} \left\{ \left| E^L_{L_\pi,J}\right|^2 + 
\left| M^L_{L_\pi,J}\right|^2 \right\}~.
\eeqa
Note that there are no electric times magnetic multipole  interference
terms due to the selection rules given above. Similarly, the 
longitudinal cross section is given entirely in terms of the
longitudinal multipoles
\beqa
\sigma_L &=& {|\vec{q \,}| \over |\vec{k \,}|} \,
\frac{1}{3} \sum_{m',m}\, \int d\Omega\, 
\left| t_{m',0,m} (\theta) \right|^2 \nonumber \\
&=& 4\pi \, {|\vec{q \,}| \over |\vec{k \,}|} \, \frac{8}{3} \,
{-k^2 \over k_0^2} \sum_{L_\pi,L,J} \,  \left| L^L_{L_\pi,J}
\right|^2~.
\eeqa
At threshold, only the three multipoles $E_{01}^1$, 
$M_{01}^2$ and $L_{01}^1$ contribute and these define the
transverse and longitudinal S--wave cross section $a_{0d}$ 
as used in \cite{BKM}, 
\beq
a_{0d} = |E_d|^2 + \varepsilon_L \,  |L_d|^2~,
\eeq
with $|E_d|^2 \equiv  |E_{01}^1|^2 + |M_{01}^2|^2$ and 
$|L_d|^2 \equiv  |L_{01}^1|^2$.
This concludes the formalism needed in this paper.
 
\section{Effective field theory}
\def\theequation{\arabic{section}.\arabic{equation}}
\setcounter{equation}{0}
\label{sec:EFT}

In this section, we briefly discuss the effective chiral Lagrangian
underlying our calculations and the corresponding power counting.
For previous related work on pion photoproduction off nuclei see
\cite{BLM,BBLMvK}.

\medskip\noindent
At low energies, the relevant degrees of freedom are hadrons, in 
particular the Goldstone bosons linked to the spontaneous symmetry
violation. We consider here the two flavor case and thus deal with the
triplet of pions, collected in the matrix $U(x)$.
It is straightforward to build an effective Lagrangian
to describe their interactions, called ${\cal L}_{\pi\pi}$. This Lagrangian
admits a dual expansion in small (external) momenta and quark (meson) 
masses as detailed below. Matter fields such as nucleons can also be 
included in the effective field theory based on the
familiar notions of non--linearly realized chiral symmetry.
These pertinent effective Lagrangian splits into two parts,
${\cal L}_{\pi N}$ and ${\cal L}_{NN}$, with the first (second) one
consisting of terms with  exactly one (two) nucleon(s) in the initial 
and the final state. Terms with more nucleon fields are of no relevance
to our calculation. The pertinent contributions to neutral pion 
photoproduction at threshold are organized according to the standard
power counting rules, which for a generic matrix element involving the
interaction of any number of pions and nucleons can then be written in
the form
\begin{equation}
{\cal M}={q^\nu}{\cal F}(q/\mu),
\end{equation}
where $\mu$ is a renormalization scale, and $\nu$ is a counting index, i.e.
the chiral dimension of any Feynman graph. $\nu$ is, of course, intimately
connected to the chiral dimension $d_i$ which orders the various terms in the
underlying effective Lagrangian (for details, see \cite{BKMrev}). 
For processes with the same number of nucleon lines
in the initial and final state $(A)$, one finds \cite{weinnp}
\begin{eqnarray}
\nu &=&4-{A}-2C+2L+{\sum _i}{V_i}{\Delta _i}\nonumber \\
& &{\Delta _i}\equiv {d_i}+{n_i}/2-2.
\label{index}
\end{eqnarray}
where $L$ is the number of loops,
$V_i$ is the number of vertices of type $i$, $d_i$ is
the number of derivatives or powers of $M_\pi$ which contribute to an
interaction of type $i$ with $n_i$ nucleon fields, and $C$ is the number of
separately connected pieces.
This formula is important because chiral symmetry places a lower
bound: ${\Delta _i}\geq 0$. Hence the leading {\it irreducible} graphs
are tree graphs ($L=0$) with the maximum number $C$ of separately
connected pieces, constructed from vertices with $\Delta _i =0$.
In the presence of an electromagnetic field, this formula is
slightly modified. Photons couple via the electromagnetic
field strength tensor and by minimal substitution.
This has the simple effect of modifying the lower bound on $\Delta _i$ to
${\Delta _i}\geq -1$, and of introducing an expansion in
the electromagnetic coupling, $e$. Throughout, we work to first order in $e$,
with one exception to be discussed below. For more details on the
counting, we refer to \cite{BLM}.
In what follows, we will work within the one--loop approximation to order
$q^3$ (notice that we refer here to the chiral dimension used to organize
the various terms in the calculation of the single--nucleon 
photoproduction amplitudes), with the exception of the S--wave
contribution to the elementary process $\gamma^\star N \to \pi^0 N$
(as discussed in the introduction).
In terms of the counting index $\nu$, we include all terms
with $\nu = 4-3A=-2$ and   $\nu = 5-3A=-1$.
Consequently, the effective Lagrangian consists of the following pieces:
\begin{equation}\label{eq:Leff}
{\cal L}_{\rm eff} = {\cal L}_{\pi\pi}^{(2)} + {\cal L}_{\pi N}^{(1)}
 + {\cal L}_{\pi N}^{(2)}  + {\cal L}_{\pi N}^{(3)}
 [+ {\cal L}_{\pi N}^{(4)}]  + {\cal L}_{N N}^{(0)}  + {\cal L}_{N N}^{(2)} \,\, ,
\end{equation}
where the index $(i)$ gives the chiral dimension $d_i$ (number of derivative
and/or meson mass insertions).
The form of ${\cal L}_{\pi\pi}^{(2)}+ {\cal L}_{\pi N}^{(1)} $ is standard.
The terms from  ${\cal L}_{\pi N}^{(3)} + {\cal L}_{\pi N}^{(4)}$ 
contributing to the single--nucleon electroproduction amplitudes 
are given in Ref.\cite{BKMe}. Note that the square brackets in
Eq.~(\ref{eq:Leff}) indicate that such fourth order terms are only
taken for the S--wave single nucleon production amplitudes\footnote{In
  fact, one also has to account for one particular dimension five
  operator as explained in \cite{BKMe}, see also Sect.~\ref{sec:ss}.}.
The effective Lagrangian can also be used to generate deuteron
wave functions of sufficient precision, as done in
Refs.~\cite{EGME,Eetal} based on a modified Weinberg power counting. We use
the NLO and the node--less NNLO* wave functions from \cite{Eetal} for
the allowed cut-off range $\Lambda = 500 \ldots 600\,$MeV, where
the cut--off stems from the regulator function in the
Lippmann--Schwinger equation used to generate the bound and the
scattering states (for a more detailed discussion giving also
the explicit form of ${\cal L}_{N N}^{(0,2)}$, see e.g.
Ref.~\cite{EGMII}). As a check,  we have also made  use of the hybrid
approach of \cite{Weinhyb}, sewing precise phenomenological wave
functions to the chirally expanded kernel. None of the results 
discussed later depend on the choice of wave function and we therefore
will only present numbers for the chiral EFT wave functions.
After these general remarks, let us now turn to the actual calculations.

\section{Anatomy of the calculation}
\def\theequation{\arabic{section}.\arabic{equation}}
\setcounter{equation}{0}
\label{sec:anat}

In this section, we outline how the various contributions to the
multipoles and the observables  are calculated. First, we briefly
discuss the separation of the transition matrix elements 
into two-- and three--body terms (or, in nuclear physics language,
impulse and meson--exchange terms). Then, these two types of contributions
are discussed separately, in particular we stress the differences to the
threshold calculation of \cite{BKM}. Many details are relegated to the appendices.
 
\subsection{General remarks}
\label{sec:gen}
\begin{figure}[htb]
   \vspace{0.5cm}
   \epsfysize=3.8cm
   \centerline{\epsffile{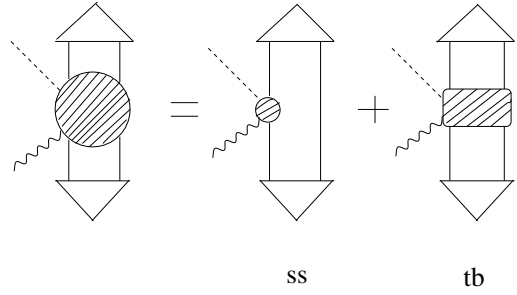}}
   \centerline{\parbox{11cm}{\caption{\label{fig:sstb}
Decomposition of the full interaction kernel (as shown in leftmost
diagram by the shaded circle) into the single scattering (ss) and 
the three--body (tb) contribution. The triangle symbolizes the 
deuteron wave function.
  }}}
\end{figure}

Consider first a generic diagram for neutral pion electroproduction
off the deuteron as shown in Fig.~\ref{fig:sstb}. The interaction
kernel decomposes into two distinct and different pieces. First,
the virtual photon can produce the pion on either the proton of the
neutron, with the other nucleon acting as a mere spectator. This
is called the single scattering contribution (ss), compare 
Fig.~\ref{fig:sstb}. It is important to note that to properly account
for this one has to transform from the photon--nucleon to the
photon--deuteron center--of--mass system as discussed below. Second,
all other terms in the interaction kernel involve both nucleons,
comprising the so--called three--body (tb) interactions (see again
Fig.~\ref{fig:sstb}). 
To be specific, the transition matrix element for pion production of the
deuteron has the form
\beq
{\cal M}^\lambda = 
\, \sum_{m_s,m_s'} |\, m_s'\rangle \langle \vec{p}_d\,' \, \vec{q} \, | a_\mu^\lambda 
J_d^\mu \, | \vec{p}_d \, \vec{k} \, \rangle \langle m_s \,|
\eeq
in terms of the three--momenta of the in--coming (out--going) deuteron,
$\vec{p}_d$ and $\vec{p}_d\,'$, respectively.  
Here, $J_d^\mu$ denotes the vector current impinging
on the deuteron. To unravel the underlying structure of the deuteron,
one expresses this matrix element in terms of the two--nucleon
current, $J_{\rm NN}$. This current then  separates into the two
terms just discussed,
\beqa
 \langle \vec{p}_d\,' \, \vec{q} \, | a_\mu^\lambda J_d^\mu \, | 
\vec{p}_d \, \vec{k} \, \rangle &=& {1 \over (2\pi)^3} \sqrt{ {E_d E_d'
\over 4 E_1 E_1' E_2 E_2' }} \,  \langle \vec{p}_d\,' \, \vec{q} \, | 
a_\mu^\lambda J_{NN}^\mu \, | \vec{p}_d \, \vec{k} \, \rangle \nonumber \\
&=&  {1 \over (2\pi)^3} \sqrt{ {E_d E_d'\over 4 E_1 E_1' E_2 E_2' }} \,
\left(  \langle \vec{p}_d\,' \, \vec{q} \, |a_\mu^\lambda J_{\rm NN}^\mu \, | 
\vec{p}_d \, \vec{k} \, \rangle^{\rm ss} +
 \langle \vec{p}_d' \, \vec{q} \, |a_\mu^\lambda J_{\rm NN}^\mu \, | 
\vec{p}_d \, \vec{k} \, \rangle^{\rm tb} \right)~,
\eeqa
where $E_i \, (E_i')$ denotes the energy of nucleon $i$ ($i =1,2$) in
the initial (final) state. In what follows, we will work
in the approximation that pion is produced in S-- or P--waves
only, that is we allow for $L_\pi = 0,1$. This means that we have
to consider three S--wave ($E_{01}^1, L_{01}^1, M_{01}^2$) and
eight P--wave multipoles ($E_{11}^2, E_{12}^2, M_{10}^1, M_{11}^1,
M_{12}^1, M_{12}^3, L_{11}^0$, $L_{11}^2, L_{12}^2$), compare Table~\ref{tab:val}. 
Only if one allows for higher partial waves,
all thirteen different structures discussed in Section~\ref{sec:mult}
will contribute.  We will now discuss the single scattering and the 
three--body contributions in more detail.

\subsection{Single scattering contribution}
\label{sec:ss}
In the previous paragraph, we have introduced the single scattering
contribution to the two--nucleon current. It has the following generic
form
\beq
a_\mu^\lambda \, J_{\rm NN}^{{\rm ss}, \mu} = 
\frac{1}{2} \left( 2m (2\pi)^3 \, \delta(\vec{p}_1\,' - \vec{p}_1 ) \,
\left[ a_\mu^\lambda \, J ^{\pi^0 p, \mu} +  a_\mu^\lambda \,
J ^{\pi^0 n, \mu} \right]_2 + ( 1 \leftrightarrow 2) \right) ~,
\eeq
with $\vec{p}_1 - \vec{p}_1\,' = \vec{p} - \vec{p}\,' - \vec{k}/2 +
\vec{q}/2$ in terms of the in-coming (out-going) relative nucleon cms
momenta $\vec{p}\, (\vec{p}\,')$, the pion and the photon
momenta and similarly for $\vec{p}_2 - \vec{p}_2\,'$. The isoscalar
current is then expressed in terms of the conventional CGLN \cite{CGLN}
amplitudes,
\beq
\left[ a_\mu^\lambda \, J ^{\pi^0 p, \mu} +  a_\mu^\lambda \,
J ^{\pi^0 n, \mu} \right]_i = 8\pi W^\star_i \, \sum_{L_\pi=0}^\infty
\tilde{O}_i^{\rm ss} (\hat{k}^\star, \hat{q}^\star) \cdot
\left( \begin{array}{cc} G_{L_\pi}^{\rm ss} (z^\star) & 0 \\
               0 & H_{L_\pi}^{\rm ss} (z^\star) \end{array}\right)
\cdot \left( \tilde{M}_{L_\pi}^{{\rm ss}, \pi^0 p} + 
 \tilde{M}_{L_\pi}^{{\rm ss}, \pi^0 n} \right)~,
\eeq
where the stared quantities refer to the center--of--mass system
of nucleon $i$, the $\tilde{M}_{L_\pi}^{{\rm ss}}$ denote the
single nucleon multipoles, the explicit form of the operators 
$\tilde{O}_i^{\rm ss}$ is given in appendix~\ref{app:cmss} and
the corresponding transformation matrix for the transverse and
longitudinal multipoles is standard \cite{BDW}. Of course, one
has to transform these expressions from the pion--nucleon to the
pion--deuteron center--of--mass system. This is described in some
detail in appendix~\ref{app:cmss}. When sandwiched between the deuteron
wave functions, one has to deal with integrals of the type
\beq\label{overlap}
\int d^3 p \, \phi^* \left(\vec{p}\,-\frac{\vec{k}}{2}+
\frac{\vec{q}}{2}\right) \, \left( {\cal
    O}_1 + \vec{{\cal O}}_2 \cdot \vec{p}\, \right) \, 
\phi \left(\vec{p}\,\right)~,
\eeq
where $\phi (p) $ denotes the momentum space deuteron wave function,
and ${\cal O}_1$, $\vec{{\cal O}}_2$ are arbitrary spin structures.
To work in coordinate space, one has to Fourier transform these
expressions, using the basic integrals
\beqa
\int d^3 p \, \phi^* \left(\vec{p}\,-\frac{\vec{k}}{2}+
\frac{\vec{q}}{2}\right) \,  {\cal O}_1  \, 
\phi \left(\vec{p}\,\right) &=&
\int d^3 r \, \phi^\dagger (\vec{r} \,) \,  {\cal O}_1  \, 
 \phi (\vec{r} \,) \, {\rm e}^{-i\left( \vec{k}-\vec{q}\,
\right) \cdot \vec{r}/2}  ~, \nonumber \\
\int d^3 p \, \phi^* \left(\vec{p}\,-\frac{\vec{k}}{2}+
\frac{\vec{q}}{2}\right) \,  {\cal O}_{2,i} \, p_i \, 
\phi \left(\vec{p}\,\right) &=&
\int d^3 r \, \phi^\dagger (\vec{r} \,) \,  {\cal O}_{2,i}  \,
\frac{1}{i} \, {\partial_r}_i \,  
 \phi (\vec{r} \,) \, {\rm e}^{-i\left( \vec{k}-\vec{q}\,
\right) \cdot \vec{r}/2}  ~,
\eeqa
and further decomposing the deuteron co--ordinate space wave function
$\phi(\vec{r} \,) $ into its radial S-- and D--wave components,
\beq
\phi(r) = \frac{1}{\sqrt{4\pi}} \left( {u(r) \over r} + \sqrt{1\over 8}
{w(r) \over r}\, S_{12} (\hat{r}) \right)~.
\eeq
Here, $S_{12} (\hat{r})$ is the usual second order tensor operator.
We also remark that in \cite{BKM} we had expressed the overlap
integrals in a factorized form, more specifically, the
transverse and longitudinal deuteron multipoles could be written
as products of the single nucleon S--wave multipoles and  a set of 
deuteron form factors. Such a procedure becomes very complicated above
threshold and is not transparent, therefore we do not follow such a
path here in detail (although we have performed some calculations
in that framework to have a further check on the numerics). 
Still, it is important to stress that the single scattering
contribution is strongly suppressed with increasing photon virtuality
because of the decreasing overlap integrals in Eq.~(\ref{overlap}).

\medskip\noindent
Again, we work in the S-- and P--wave approximation. The corresponding
single nucleon multipoles are subject to the chiral expansion. We work
to first non--trivial loop order, i.e. to third order, with the
exception of the proton and neutron  S--wave multipoles 
$E_{0+}^{n,p}$ and $L_{0+}^{n,p}$. We take the
form given in \cite{BKMe} which includes all fourth order terms 
and one particular fifth order term necessary to separate cleanly
the longitudinal from the transverse piece. We refer to that paper
for the explicit expressions of the single nucleon transition current.
Here, we only spell out the generic form for the S--waves,
\beq
{\cal S} = {\cal S}^{\rm Born} + {\cal S}^{\rm q^3-loop} +
{\cal S}^{\rm q^4-loop} + {\cal S}^{\rm ct}~,
\eeq
with ${\cal S} = E_{0+}^{p,n}$ or $L_{0+}^{p,n}$. At fourth order
one has two local operators $\sim k^2$ with the LECs $a_3^I, a_4^I$ ($I=p,n$).
However, a particular low--energy theorem (LET) strongly correlates these
two LECs, in the soft--pion limit one has $a_3^I + a_4^I =0$. To break
this correlation that is not observed in the proton data, one has to
include a correction to the LET of the form $L_{0+}^{\rm ct, I} 
= -eM_\pi^2 k^2 a_5^I$ which is formally of fifth chiral order (in the
effective Lagrangian). As noted before, with that
we are able to describe the proton data of \cite{benno,distler} for
$\gamma^\star p \to \pi^0 p$ at $k^2 = -0.1\,$GeV$^2$ but not the
more recent data of \cite{Merkel} at half the photon virtuality (as
discussed in the introduction).
For the LECs related to the neutron amplitude, we follow two strategies.
First, we fix  $a_3^n = a_4^n$ from resonance saturation 
as detailed in Ref.\cite{BKMe} and determine 
$a_5^n$ from a fit to the empirical threshold amplitude $L_d$
of Ref.~\cite{Ewald}. Second, we leave both $a_3^n$ and $a_4^n$ as free
parameters thus relaxing the constraint due to the LET and fit to the
threshold total cross sections of Ref.~\cite{Ewald}. In what follows, 
we will call these two procedures fit~1 and fit~2, respectively. The
scaling procedure performed in \cite{BKM} was done too simplistically
for the longitudinal S--wave multipole leading to the too large S--wave
cross section.

\begin{figure}[htb]
   \vspace{0.5cm}
   \epsfysize=5.5cm
   \centerline{\epsffile{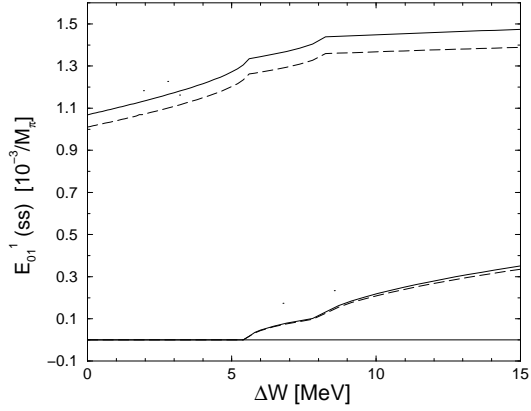}}
   \centerline{\parbox{11cm}{\caption{\label{fig:boost}
Effect of the Lorentz boost on the arguments of the multipoles. 
Shown are the real (upper two lines) and the
imaginary part (lower two lines) of the single scattering contribution
to the electric multipole $E_{01}^1$ for varying pion 
excess energy $\Delta W$ at fixed photon virtuality $k^2 =
   -0.1\,$GeV$^2$. Solid (dot-dashed) lines: with (without) boost.
  }}}
\end{figure}

\medskip\noindent
There is one additional point that deserves particular discussion.
With increasing photon excess energy (that is the energy normalized
to the threshold energy), one should observe two unitary cusps due to the
opening of the $\pi^+nn $ and the $\pi^- pp$ channel for neutral
pion production off the proton and the neutron, respectively.
Indeed, if one does not boost the energy dependent multipoles as
suggested in \cite{KW}, two cusps are visible  as shown in Fig.~\ref{fig:boost}
for the real part of $E_{10}^1$ together with the corresponding 
growth of the imaginary part\footnote{Note that we work with $M_{\pi^0} =
134.97$~MeV and $M_{\pi^+} = 140.13$~MeV or $M_{\pi^+} = 142.53$~MeV, 
to account for the neutron--proton mass difference in the rescattering diagrams. A
detailed discussion of this point is given  in Ref.\cite{BKMz}.}.
The effect of applying the full boost correction, i.e. also changing the
arguments of the multipole amplitudes, turns out to be small, as
also shown in  Fig.~\ref{fig:boost}. The real part is shifted by a few
percent and the imaginary part is almost unaffected for the pion energies
considered here. Note that in all non--analytic terms,
like e.g. the energy dependence of the S--waves $\sim \sqrt{1 - 
\omega^2/\omega_c^2}$ (and similar terms in the P--waves), 
we always use the physical values for the
$nn \pi^+$ and $pp\pi^-$ thresholds given by $\omega_c$. This is consistent with the
chiral expansion and the analytic structure of the amplitudes.

\subsection{Three--body contribution}
\label{sec:tb}
\begin{figure}[H]
   \vspace{0.5cm}
   \epsfysize=5.5cm
   \centerline{\epsffile{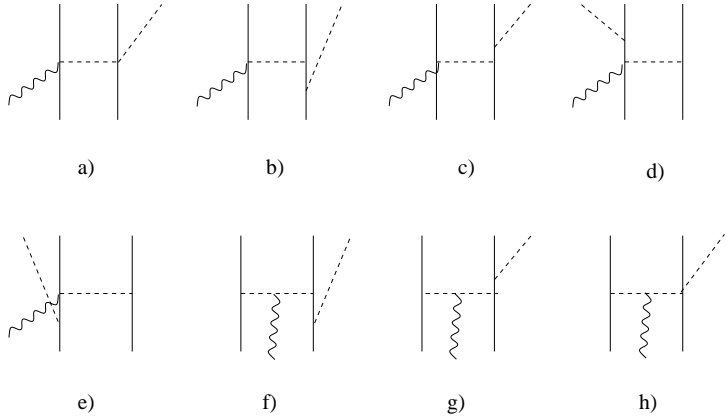}}
   \centerline{\parbox{11cm}{\caption{\label{fig:graphs3}
Three--body interactions which contribute to neutral pion electroproduction
at threshold to order $q^3$ (in the Coulomb gauge). Here, solid, dashed
and wiggly lines denote nucleons, pions and photons, in order. 
  }}}
\end{figure}

We now turn to the three--body contribution. Above threshold, we have
8 diagrams contributing at third order, see Fig.~\ref{fig:graphs3}.
Note that in contrast to earlier works, we differentiate between
the charged and neutral pion masses also in these graphs, although
this is formally an effect of higher order. We remark that one can
combine these various contributions into two distinct classes with
one and two pion propagators, respectively, the so--called
rescattering and pion-in-flight diagrams.
\noindent
The explicit analytical expressions for the corresponding matrix
elements in momentum space are:

\medskip
\noindent\underline{Diagrams a) + b) + c) + d) + e)} (rescattering type)
\beq
{\cal M}_{a+b+c+d+e} = 2 \, {e g_A m^2\over F_\pi^3} \, 
\left(  \vec{\varepsilon} \cdot \vec{S\,} \, 
\left[  {g_A^2}\, {1 \over q_0} \,  \vec{q\,} \cdot \vec{q\,}' - q_0 \right]
- g_A^2 \, {1 \over q_0} \,   \vec{S\,} \cdot {\vec{q\,}'} \, 
\vec{\varepsilon} \cdot \vec{q} \, \right) \, {1 \over
{\vec{q\,}'}\,^2  + \delta^2}\, \left[
\vec{\tau}_1 \cdot \vec{\tau}_2 - \tau_1^3 \tau_2^3 \right]~.
\eeq
Here, $q' = (q_0', \vec{q\,}' \,)$ with $q_0' = q_0 + {\cal O}(1/m)$
is the four--momentum of the exchanged pion and 
\beq
\delta^2 = M_{\pi^+}^2 - q_0^2 - i\, \epsilon~.
\eeq
\medskip
\noindent\underline{Diagrams f) + g) + h)} (pion-in-flight type)

\beq
{\cal M}_{f+g+h} = -2 \, {e g_A m^2\over F_\pi^3} \, \vec{\varepsilon}
\cdot \left(\vec{q\,}'' + \vec{q\,}'\right) \,\vec{S\,}
\cdot\vec{q\,}''
\, { {g_A^2} \, q_0^{-1} \, \vec{q\,} \cdot \vec{q\,}' - q_0 \over
({\vec{q\,}''}\,^2 + M_{\pi^+}^2) \, ({\vec{q\,}'}\,^2 + \delta^2)}\, \left[
\vec{\tau}_1 \cdot \vec{\tau}_2 - \tau_1^3 \tau_2^3 \right]~.
\eeq
Here, we use the following convention.  The intermediate pion
has momentum ${\vec q\,}''$ after emission from the left nucleon line 
and before the interaction with the photon. After that, the
momentum is ${\vec q\,}'$ and the pion is absorbed on the right nucleon.
Note that for both classes of diagrams,  the factor 2 in front
takes care of the interchange $1 \leftrightarrow 2$. We remark that
in contrast to previous work \cite{BKM,BBLMvK} we differentiate 
between the charged and neutral pion masses for the exchanged meson.
While that is formally an effect of higher order, we still consider
it here because it was already shown to be the dominant isospin breaking
effect in the investigation of pion photoproduction off nucleons,
first discussed in the context of chiral perturbation theory in \cite{BKMpiN}.
In essence, we have calculated all three-body graphs for the two different
values of  $\omega_c$ corresponding to the opening of the $pp \pi^-$ 
and the $nn\pi^+$ channels and performed the necessary average.
Since these operators are sandwiched between the deuteron wave
functions and one has to integrate over all momenta, one picks
up an imaginary part from the intermediate $NN\pi$ state, i.e.
the corresponding propagators have to be split into a real (principal
value) and an imaginary part. It appears when the momentum
of the exchanged pion is equal to the charged pion mass, that is
at an excess energy of 
\beq
\Delta W_c = \omega_c + \sqrt{m_d^2+\omega_c^2-M_{\pi^0}^2} - W_0 =
 5.3\, (7.9)\,{\rm MeV}~, 
\eeq
for the $pp \pi^-$ ($nn \pi^+)$ intermediate state. Here,
$W_0 = m_d + M_{\pi^0}$ is the threshold energy. These poles will reveal
itself  as a cusp--like structure in the corresponding multipoles (if one
considers the three--body terms separately, see Section~\ref{sec:res}).
As before, we have restricted the calculation to relative
S-- and P--waves. We have also performed calculations
without this truncation, which gives a relative measure of the contribution from
D-- (and higher partial) waves. More specifically, consider a typical coordinate
space integral,
\beq\label{eq:full}
\int d\Omega_r \, {\rm e}^{-i \left( \vec{k} x + \vec{q}/2 \right)\cdot \vec{r}}
= 4\pi j_0 (a r) = 4 \pi \sum_{L=0}^\infty (-{\rm sgn}(x))^L (2L+1) \,
j_L (b) j_L (c) \, P_L (\hat{q}\cdot \hat{k})~,
\eeq
with $a = |\vec{k} x + \vec{q}/2| \,r$, $b = kr |x|$ and $c = qr/2$. Note that
the coefficient $a$ in the Bessel function $j_0 (ar)$ contains the 
explicit angular dependence. In case of the  S-- and P--wave approximation,
one  operates with the projection matrix on the full sum and the series in
Eq.~(\ref{eq:full}) truncates after the first few terms.
We have found that these differences are
very small, justifying a posteriori the assumption of only retaining the
S-- and P--waves.

\section{Results and discussion}
\def\theequation{\arabic{section}.\arabic{equation}}
\setcounter{equation}{0}
\label{sec:res}

In this section, we display the results for the multipoles, differential
and total cross sections and the S--wave cross section $a_{0d}$ for the two fit
strategies. We have performed calculations with the chiral EFT wave
functions at NLO \cite{EGMII} and NNLO* \cite{EGME,Eetal} for cut-offs
in the range of 500 to 600 MeV.  Since the results for the observables are very
similar for all these various wave functions, we only show these for
the NNLO* wave function with $\Lambda = 600\,$MeV. The
fitted LECs vary mildly  for the various wave functions as shown in 
Table~\ref{tab:LECs}. We note that all these LECs are of natural size
(note that the value for $a_5^n$ appears unnaturally large due to the
particular definition of this LEC as used in Ref.\cite{BKMe}, see also
the discussion in that paper). We remark that the results using 
the Bonn wave function as employed in Ref.~\cite{BKM} are fully consistent
with the ones based on the chiral EFT wave functions and we thus
refrain from displaying these numbers here.
\renewcommand{\arraystretch}{1.3}
\begin{table}[htb]
\begin{center}
\begin{tabular}{|c||c|c|c|c|}
\hline
w.f. & NLO-500 & NLO-600 & NNLO*-500 & NNLO*-600  \\
\hline
$a_3^n$ [GeV$^{-4}$] & \phantom{$-$}4.010 & \phantom{$-$}3.459 
                     & \phantom{$-$}4.832 & \phantom{$-$}4.966 \\ 
$a_4^n$ [GeV$^{-4}$] & $-$5.925 & $-$5.745 & $-$5.895 & $-$5.660 \\
\hline 
$a_5^n$ [GeV$^{-5}$] & $-$34.49 & $-$36.45 & $-$29.86 & $-$27.51 \\ 
\hline\end{tabular}
\parbox{12cm}{
\caption{Values of the fitted LECs for the various wave functions
(w.f.). The values for $a_{3,4}^n$ refer to the fits~2, whereas
the corresponding $a_5^n$ belongs to the respective fits~1.
\label{tab:LECs}}}
\end{center}
\end{table}

\medskip\noindent
\begin{figure}[htbp]
\psfrag{E101}{$E_{01}^1$}
\psfrag{L101}{$L_{01}^1$}
\psfrag{M201}{$M_{01}^2$}
   \vspace{0.5cm}
   \epsfysize=7cm
   \centerline{\epsffile{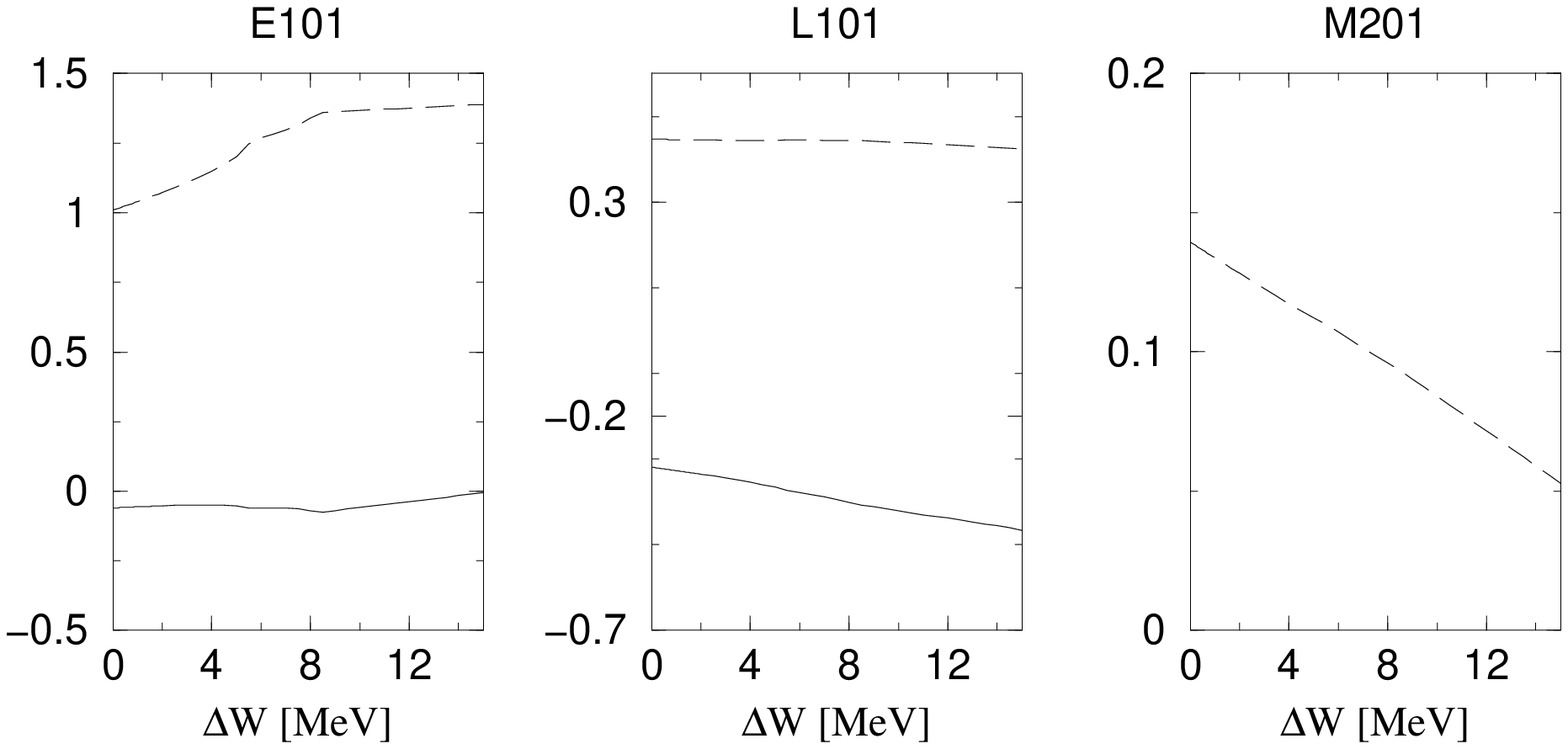}}
   \centerline{\parbox{11cm}{\caption{\label{fig:S}
S--wave multipoles as a function of the pion excess energy $\Delta W$
for photon virtuality $k^2 = -0.1\,$GeV$^2$. The solid (dashed) line
shows the total (single scattering) contribution. Note the absence of 
cusp effects in the full result for $E_{01}^1$. For $M_{01}^2$, there
is no tb contribution to this order in the chiral expansion.
Units are $10^{-3}/M_{\pi^+}$.
  }}}

\vspace{1cm}

\psfrag{E211}{$E_{11}^2$}
\psfrag{E212}{$E_{12}^2$}
   \vspace{0.5cm}
   \epsfysize=7cm
   \centerline{\epsffile{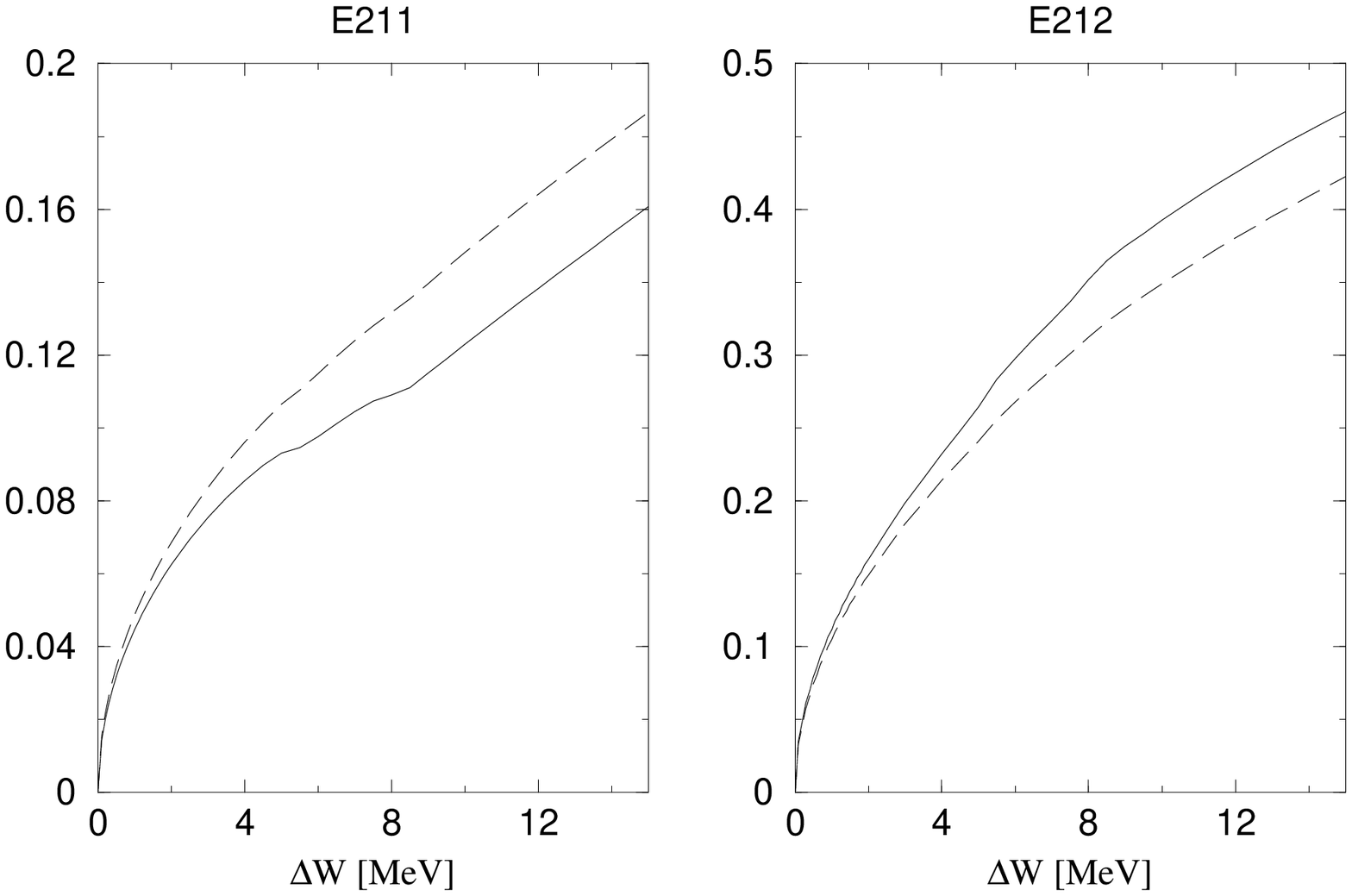}}
   \centerline{\parbox{11cm}{\caption{\label{fig:PE}
Electric P--wave multipoles as a function of the pion excess energy $\Delta W$
for photon virtuality $k^2 = -0.1\,$GeV$^2$. The solid (dashed) line
shows the total (single scattering) contribution. Units are $10^{-3}/M_{\pi^+}$.
  }}}
\end{figure}
\begin{figure}[htbp]
\psfrag{L011}{$L_{11}^0$}
\psfrag{L211}{$L_{11}^2$}
\psfrag{L212}{$L_{12}^2$}
   \vspace{0.5cm}
   \epsfysize=6cm
   \centerline{\epsffile{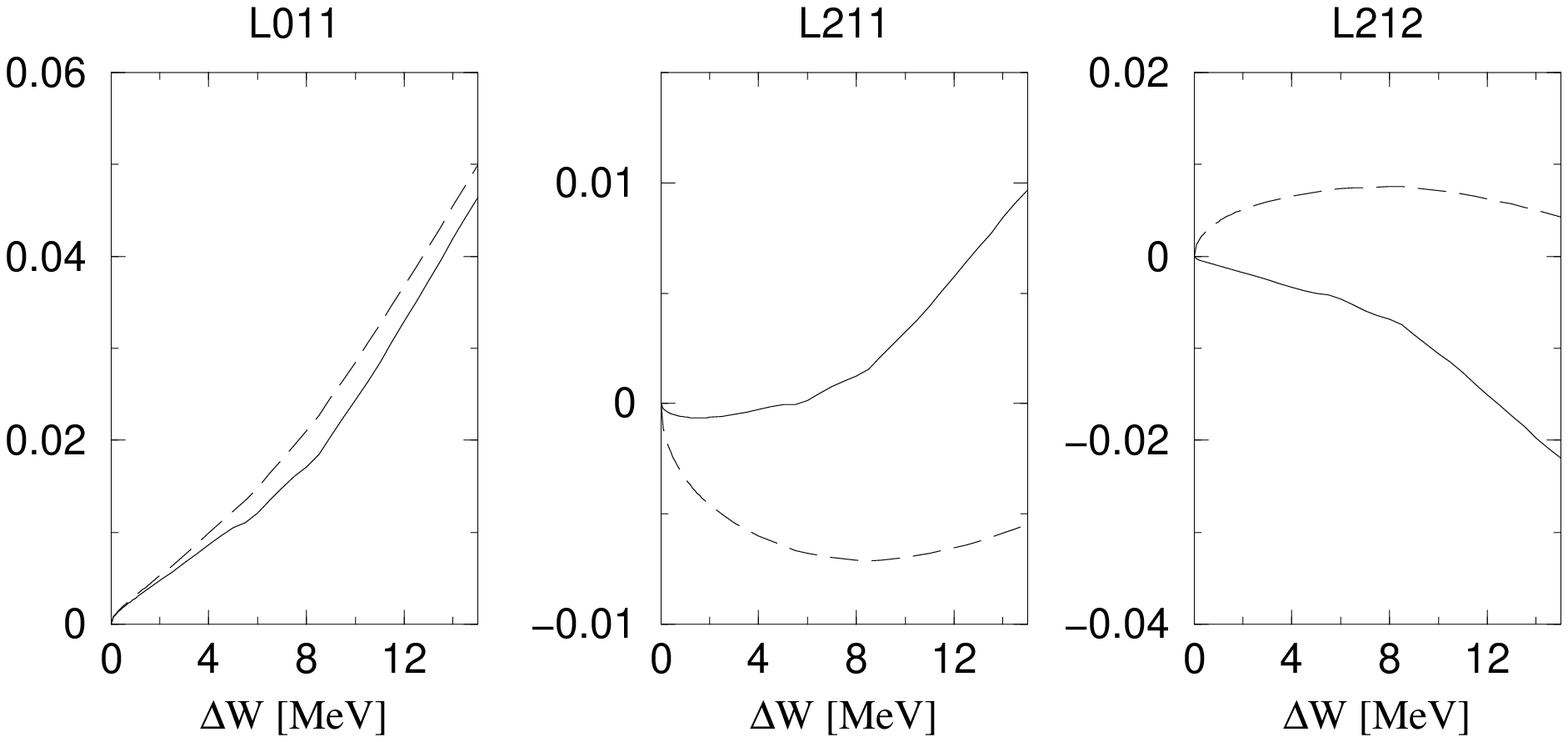}}
   \centerline{\parbox{11cm}{\caption{\label{fig:PL}
Longitudinal P--wave multipoles as a function of the pion excess energy $\Delta W$
for photon virtuality $k^2 = -0.1\,$GeV$^2$. The solid (dashed) line
shows the total (single scattering) contribution. Units are $10^{-3}/M_{\pi^+}$.
  }}}

\vspace{1cm}

\psfrag{M110}{$M_{10}^1$}
\psfrag{M111}{$M_{11}^1$}
\psfrag{M112}{$M_{12}^1$}
\psfrag{M312}{$M_{12}^3$}
   \vspace{0.5cm}
   \epsfysize=9cm
   \centerline{\epsffile{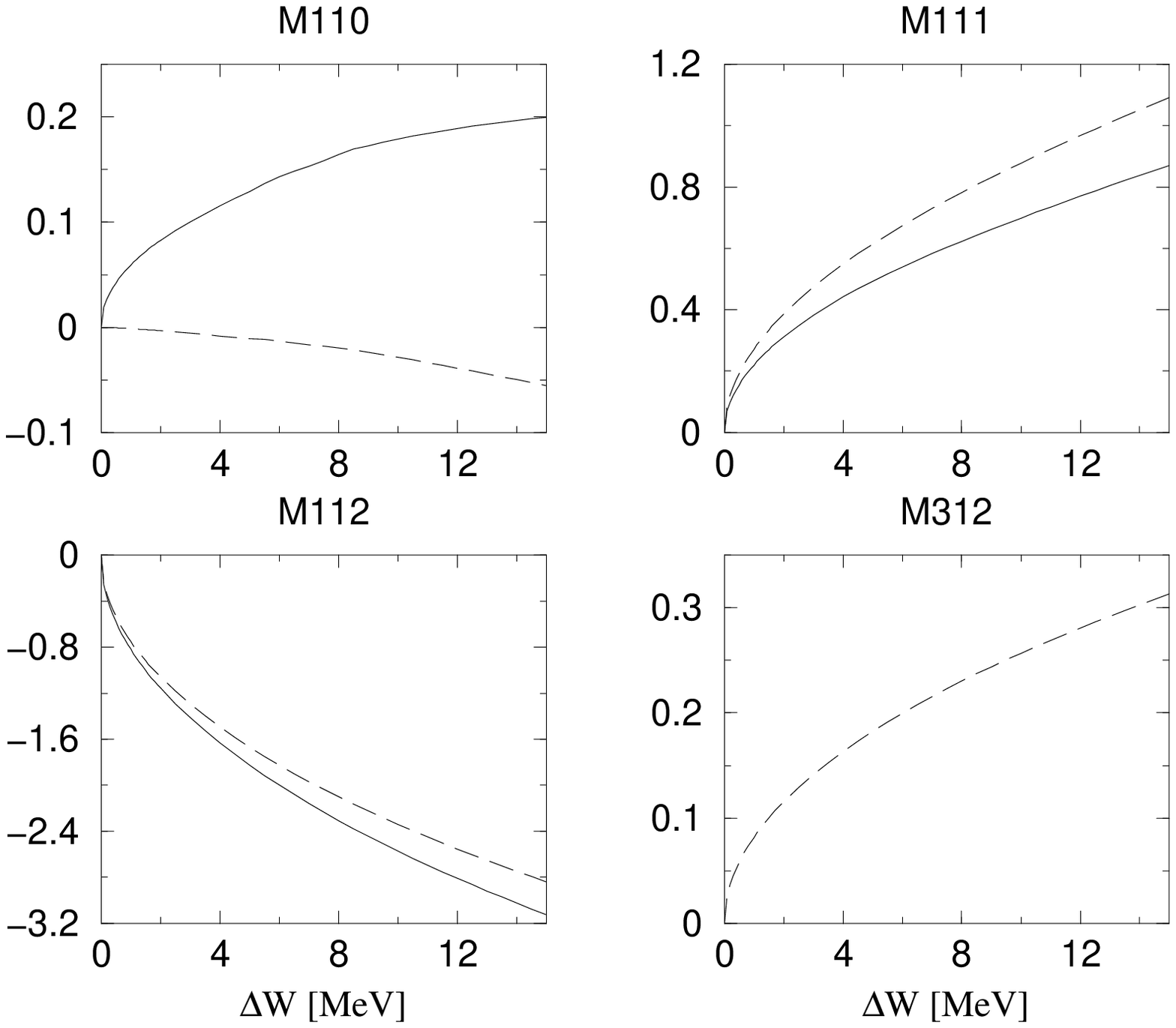}}
   \centerline{\parbox{11cm}{\caption{\label{fig:PM}
Magnetic P--wave multipoles as a function of the pion excess energy $\Delta W$
for photon virtuality $k^2 = -0.1\,$GeV$^2$. The solid (dashed) line
shows the total (single scattering) contribution.  Units are $10^{-3}/M_{\pi^+}$.
  }}}
\end{figure}
\noindent
The real parts of the various S-- and P--wave multipoles are displayed in 
Figures~\ref{fig:S}-\ref{fig:PM} by the solid lines for fit~2 using 
the NNLO* wave functions with $\Lambda = 600\,$MeV. The corresponding
single scattering contribution is also shown (dashed lines) \footnote{We refrain
from showing these multipoles  using the other wave functions for fit~2
because they come out very similar. There are some differences in the
multipoles for fit~1, as will be discussed for the S--waves later on.}. As found in
previous calculations, the three--body effects are sizeable, especially in
the S--waves. In contrast to previous attempts using meson--exchange models
this does not pose a problem for extracting the single scattering contribution
because one can systematically calculate the higher order corrections to the
three--body terms. This was indeed done for the case of neutral pion photoproduction
off deuterium in \cite{BBLMvK} and we anticipate a similar result for the case
under consideration (a complete fourth order calculation for the deuteron
case can only be done when a similar investigation for single nucleon 
electroproduction is available and the already discussed inconsistencies have
been resolved). We note in particular the  cusp--like effects in certain
P--wave multipoles due to the pion mass difference in the three--body
contributions as discussed in Section~\ref{sec:tb}. 
A comparable multipole
analysis of the data is not yet available. The multipoles extracted in
Ref.~\cite{Ewald} are based on the simplifying assumptions of constant S--waves and
P--waves that solely depend on the pion center-of-mass momentum. A direct
comparison of the multipoles obtained here with the ones of \cite{Ewald} have
therefore to be taken {\it cum grano salis}. Nonetheless we have
performed the fits of type~1 by matching to the empirical value of
$|L_d|$ to get a better handle on the theoretical uncertainties of our
calculation. Note also that the recently proposed exact  cancellation \cite{rek} 
between the single nucleon rescattering and the charge exchange three--body diagram
at threshold is visible in the S--wave multipoles, the cusp effects from the
ss and tb terms neatly cancel, cf. Fig.~\ref{fig:S}. However, this argument
only affects a subset of graphs and does not lead to the conclusion
that one is essentially sensitive to the single scattering amplitude, as reflected
in our results.

\medskip\noindent
\begin{figure}[htbp]
\epsfig{file=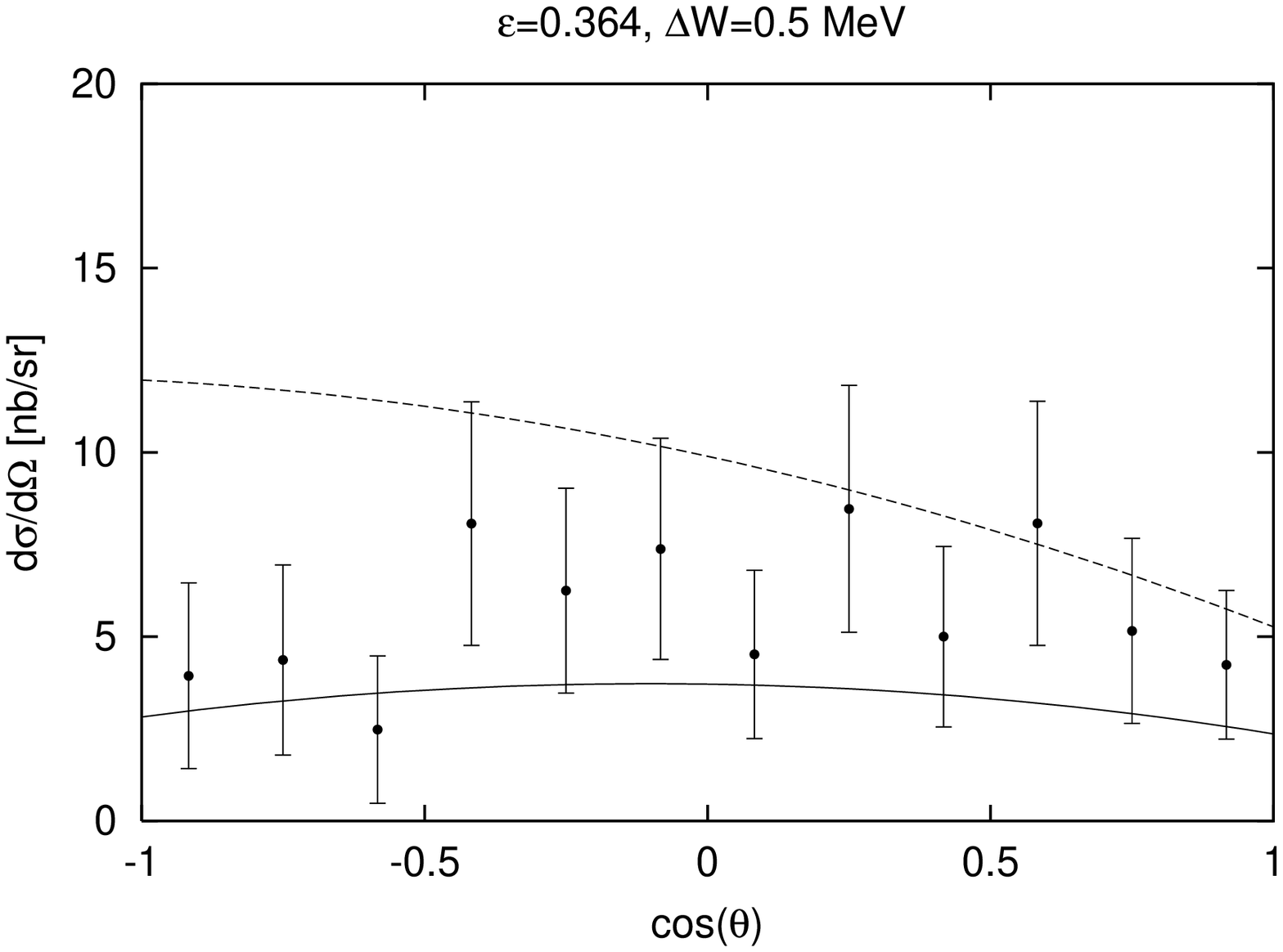,width=2.2in,angle=270}
\hfill
\epsfig{file=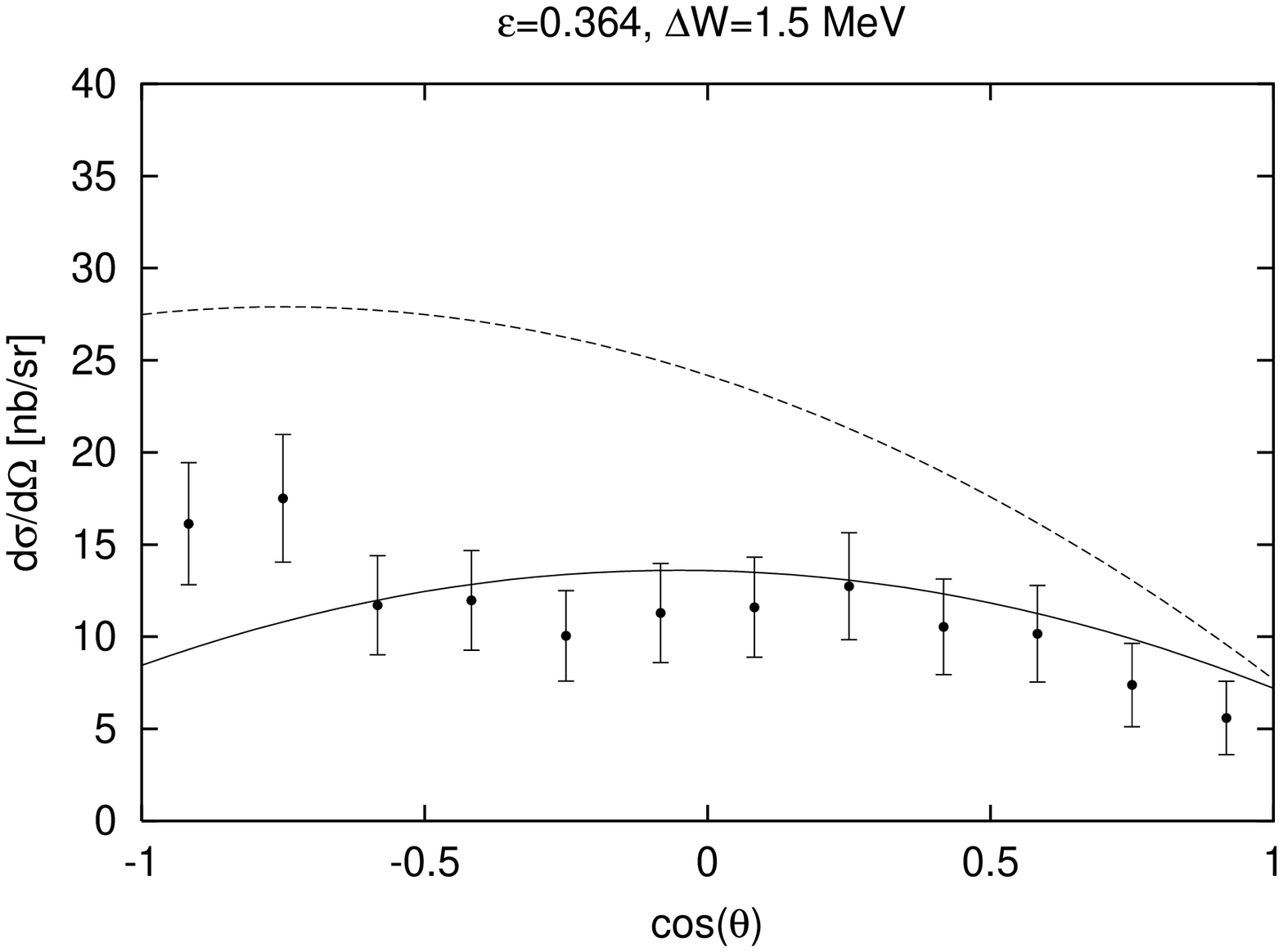,width=2.2in,angle=270}

\vspace{0.5cm}

\epsfig{file=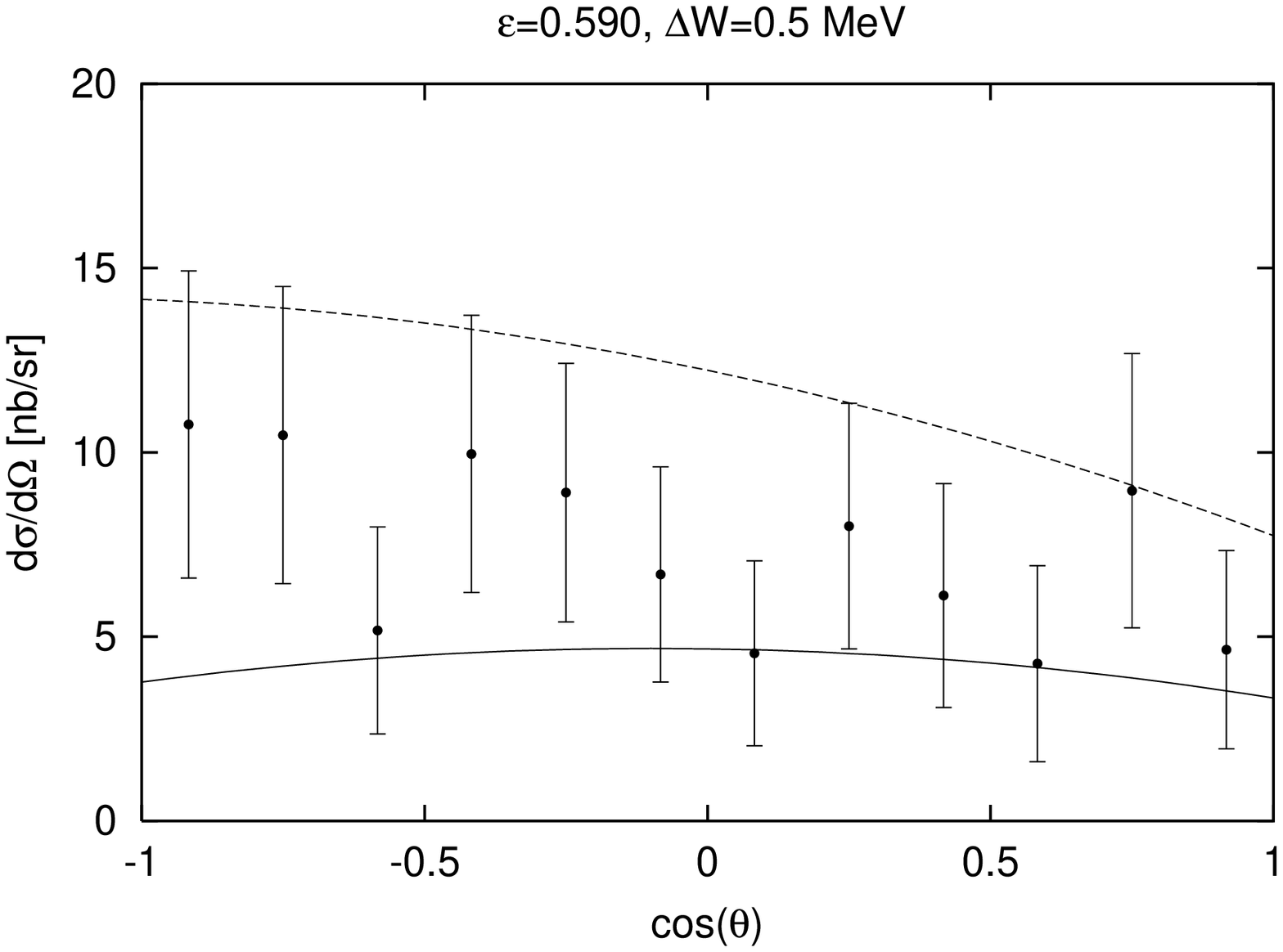,width=2.2in,angle=270}
\hfill
\epsfig{file=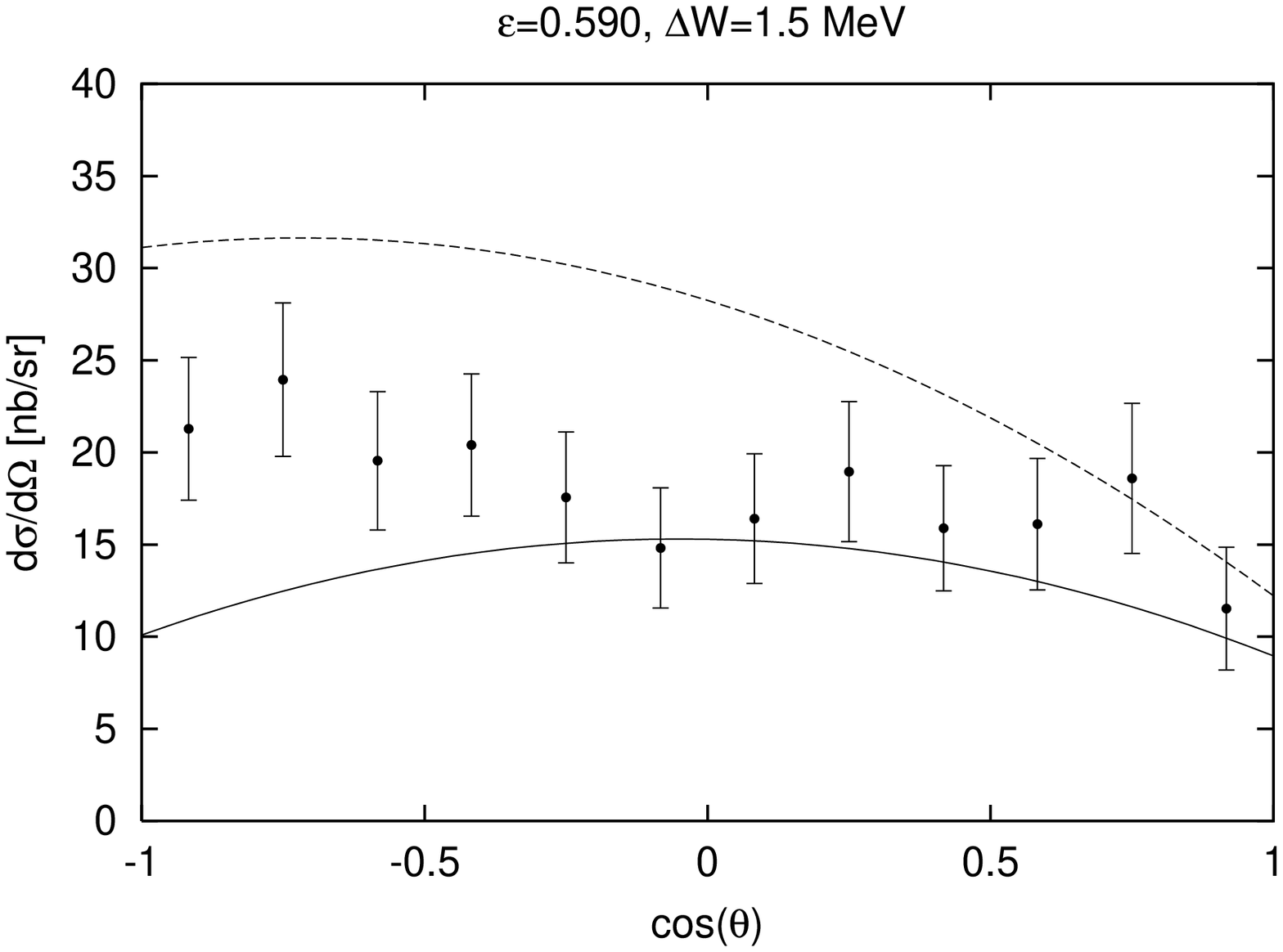,width=2.2in,angle=270}

\vspace{0.5cm}

\epsfig{file=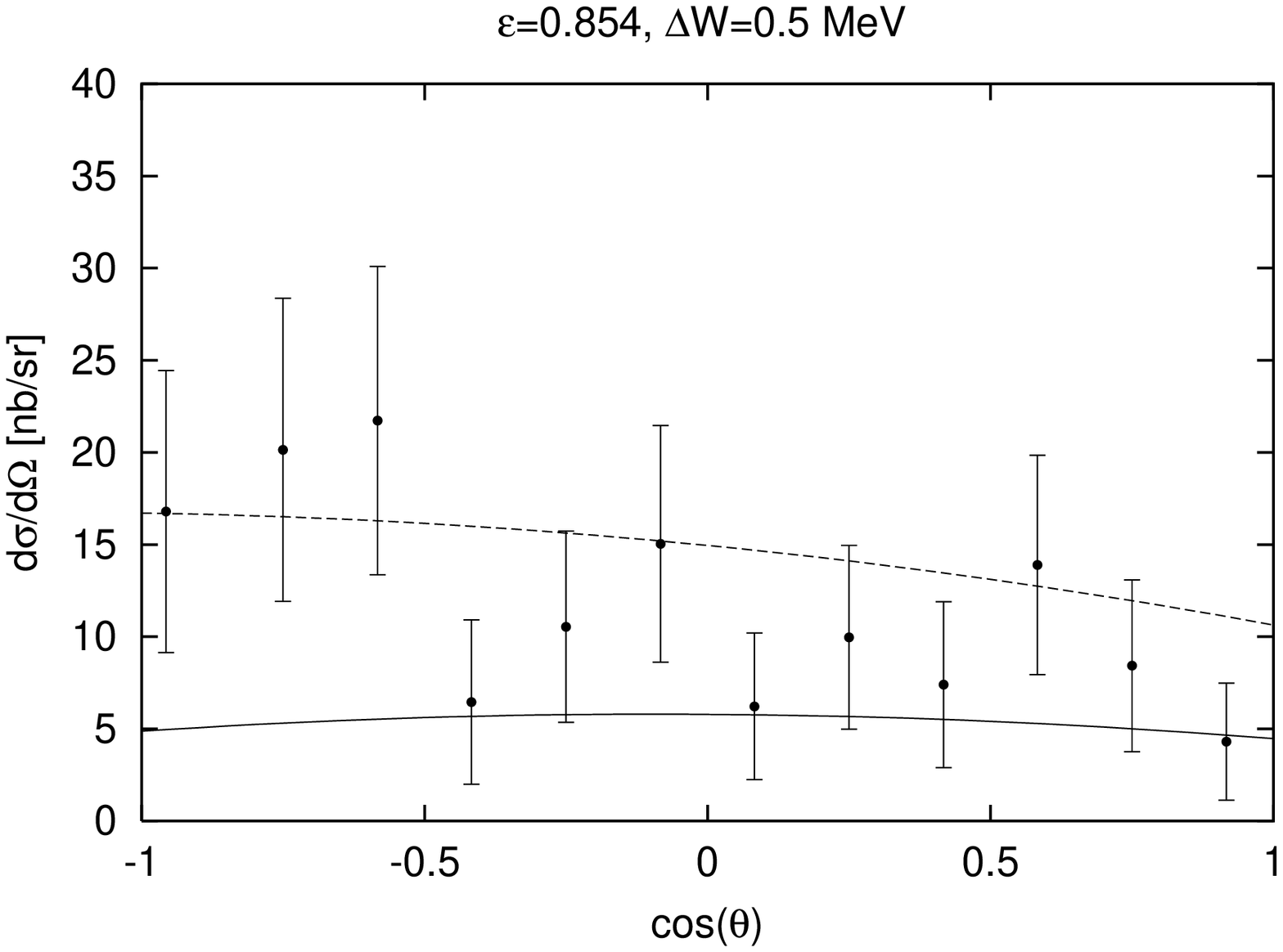,width=2.2in,angle=270}
\hfill
\epsfig{file=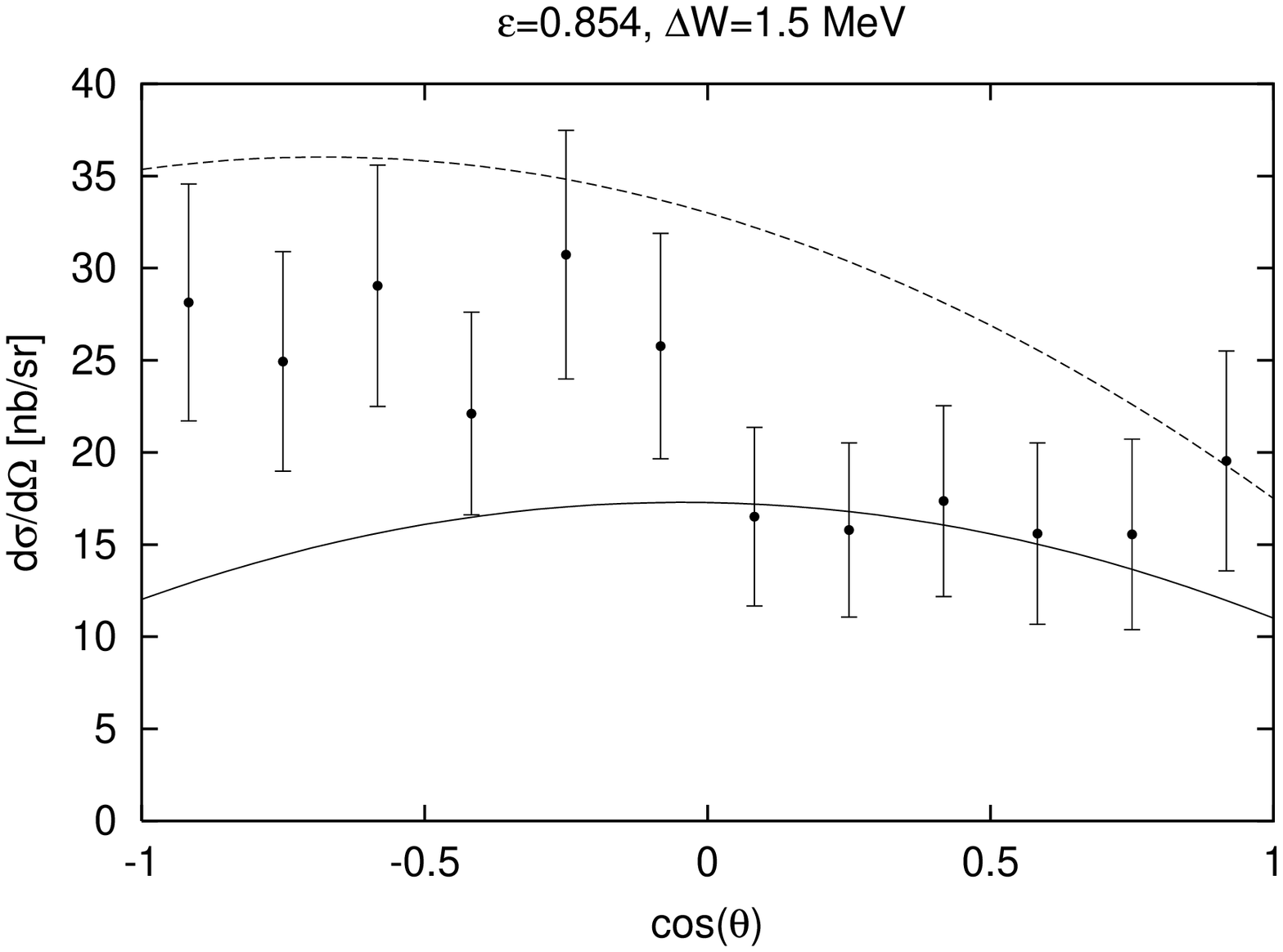,width=2.2in,angle=270}

\vspace{1cm}

\centerline{
\parbox{16cm}{\caption{Differential cross section at $\Delta W =
    0.5\,$MeV (left column) and $\Delta W = 1.5\,$MeV (right
    column) at three different values of the photon polarization
    for the NNLO*-600 wave function in comparison to the 
    MAMI data \protect\cite{Ewald}. Fit~1~(2): dashed (solid) lines.
    \label{fig:dXS0515}}}
}
\end{figure}
%
%
%
\begin{figure}[htbp]
\epsfig{file=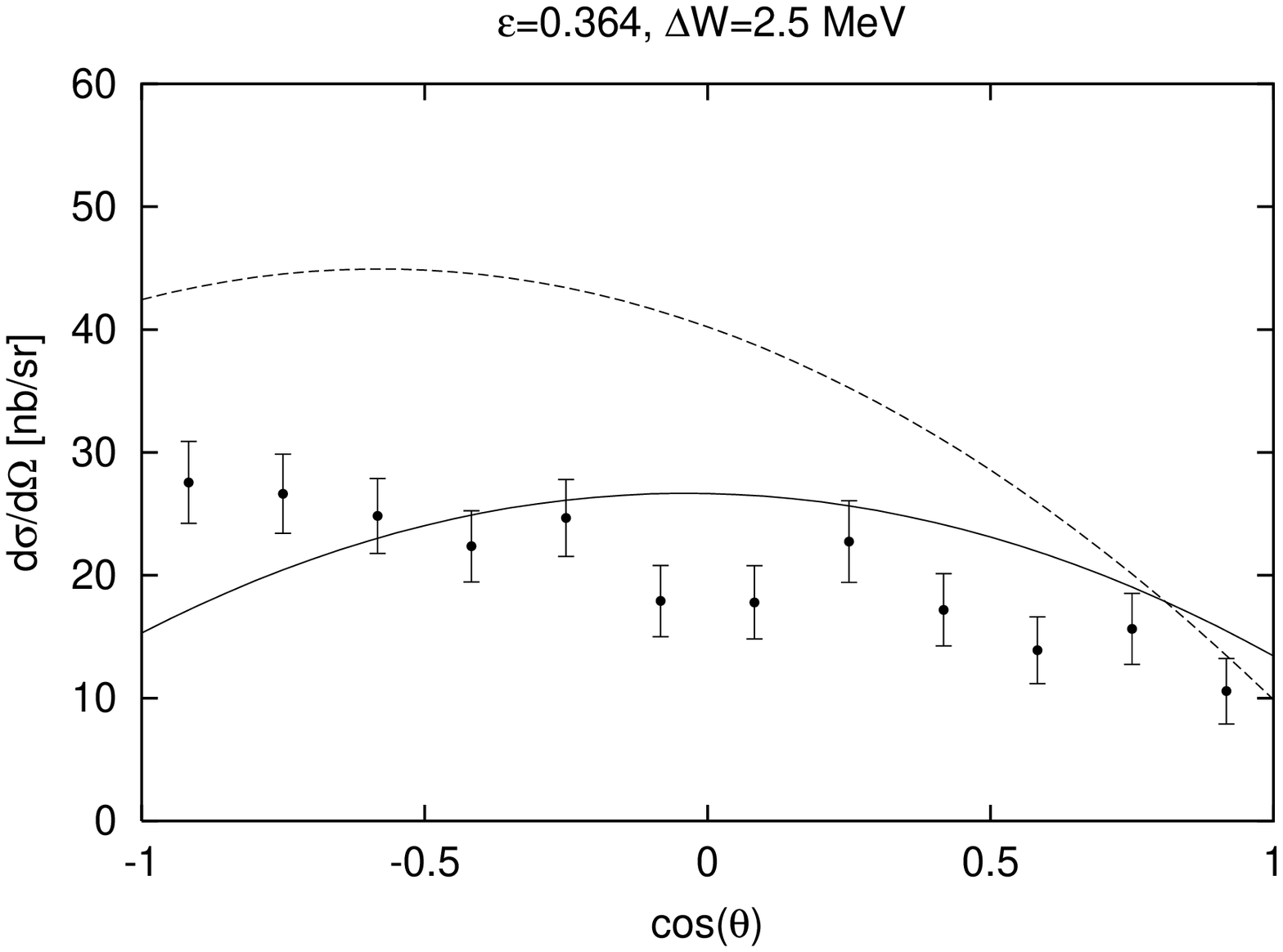,width=2.2in,angle=270}
\hfill
\epsfig{file=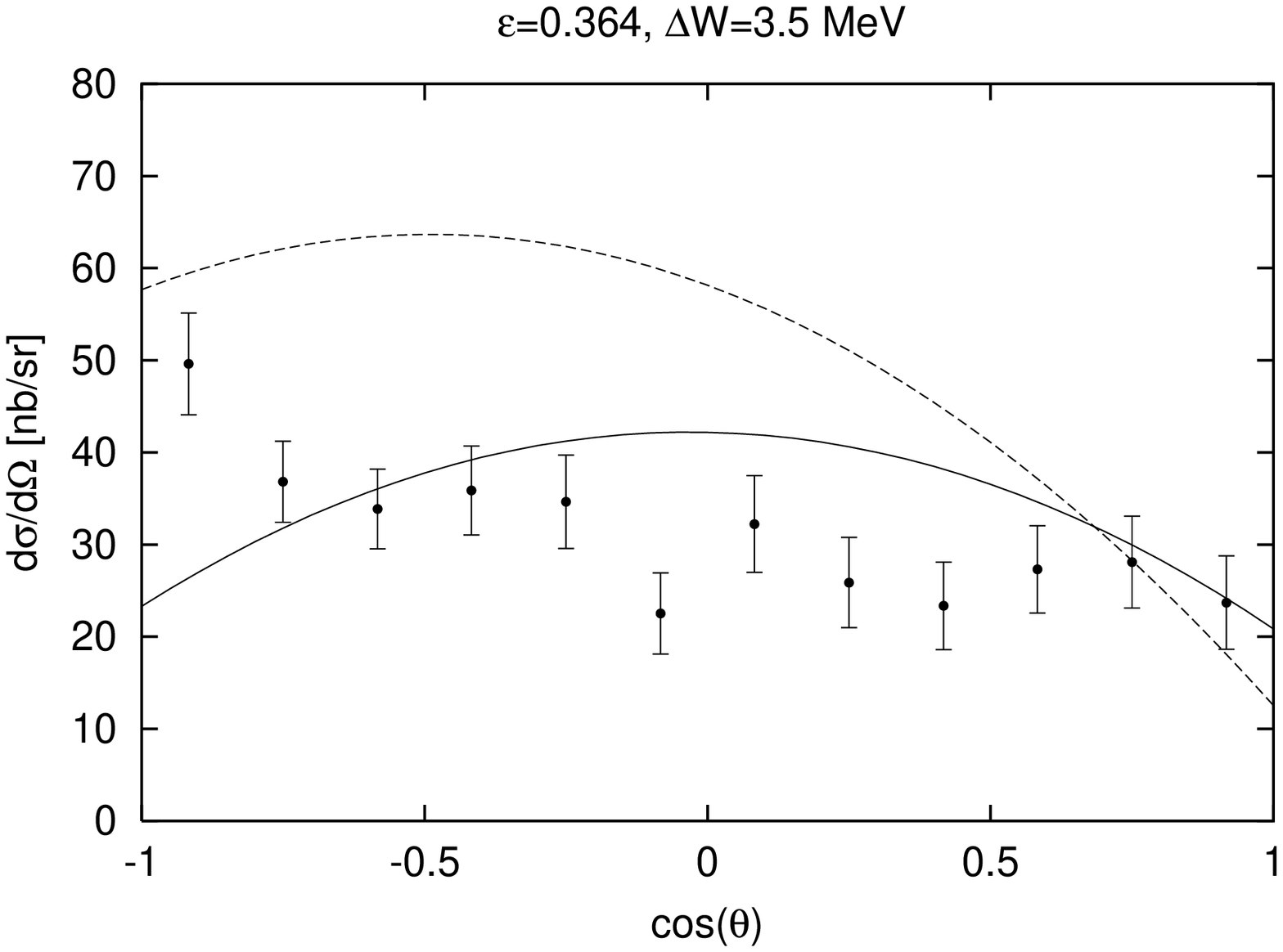,width=2.2in,angle=270}

\vspace{0.5cm}

\epsfig{file=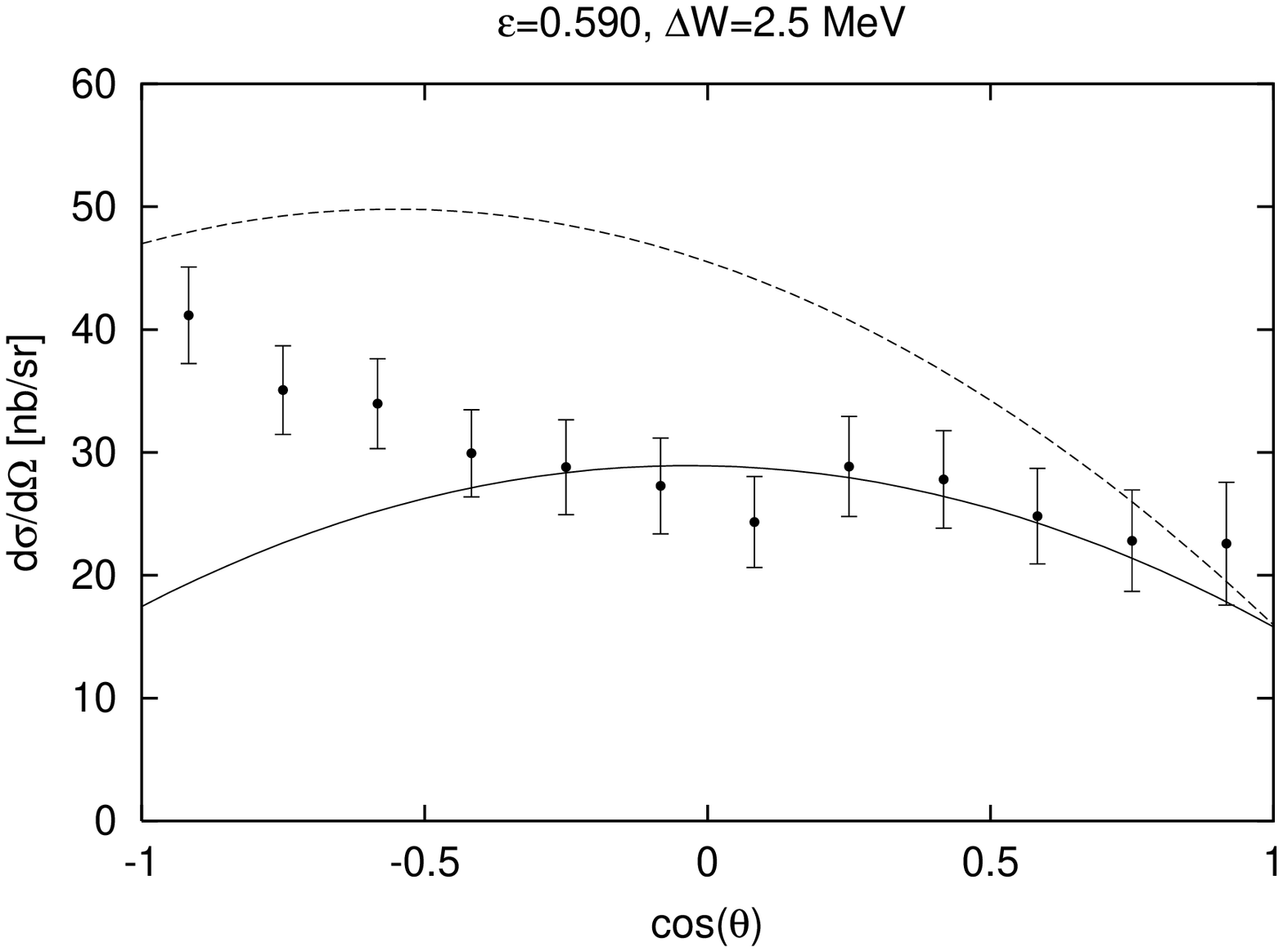,width=2.2in,angle=270}
\hfill
\epsfig{file=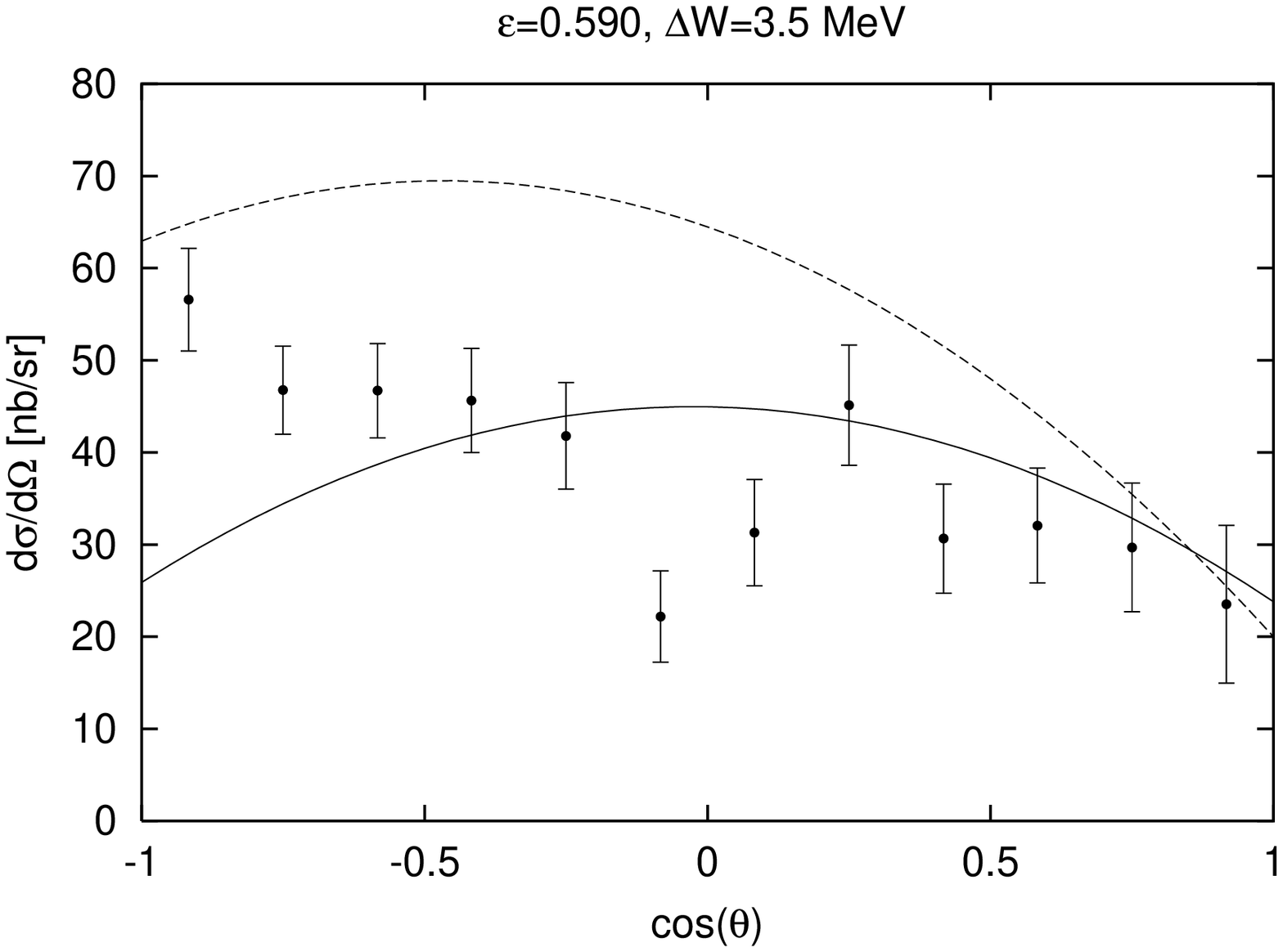,width=2.2in,angle=270}

\vspace{0.5cm}

\epsfig{file=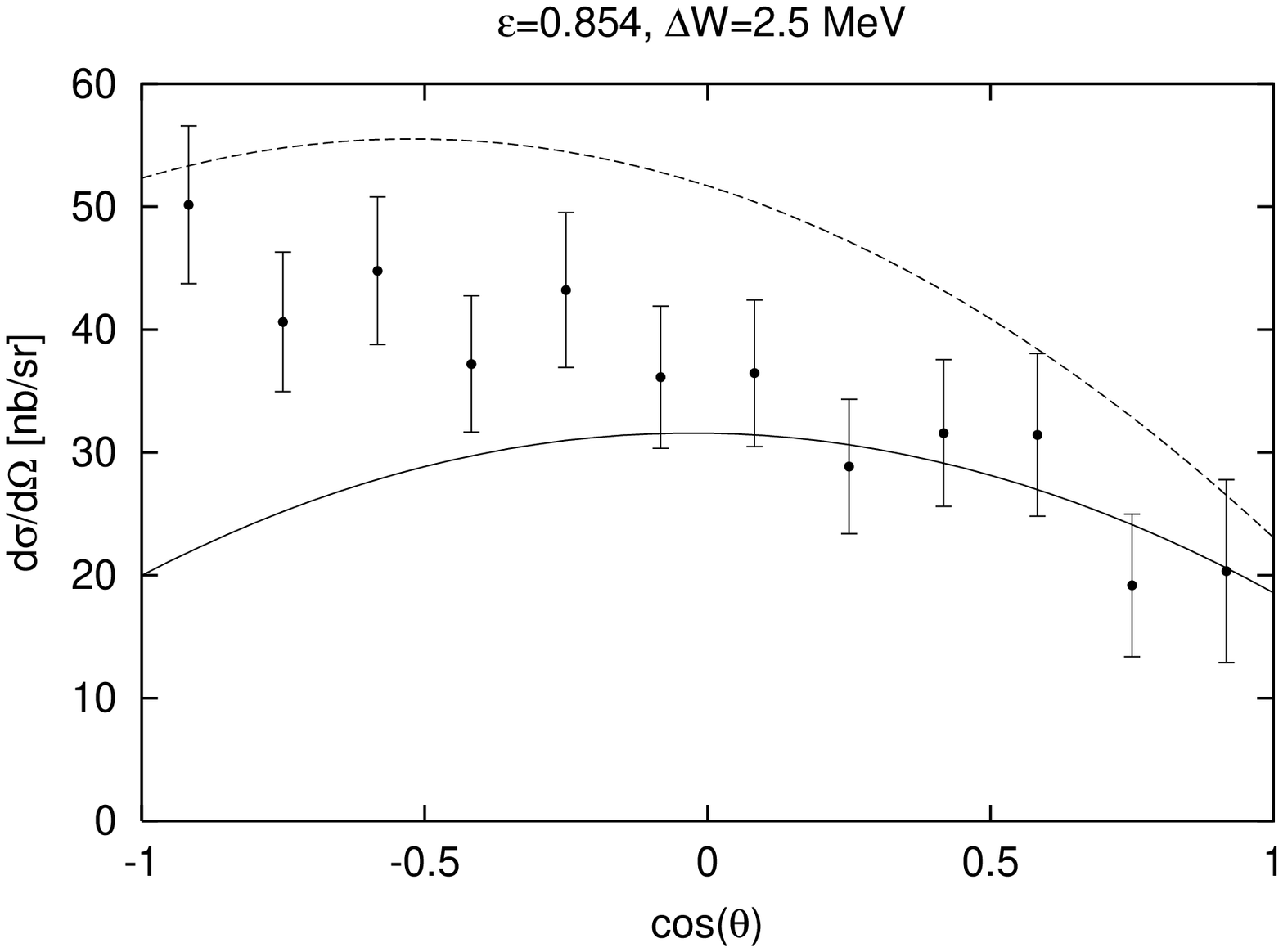,width=2.2in,angle=270}
\hfill
\epsfig{file=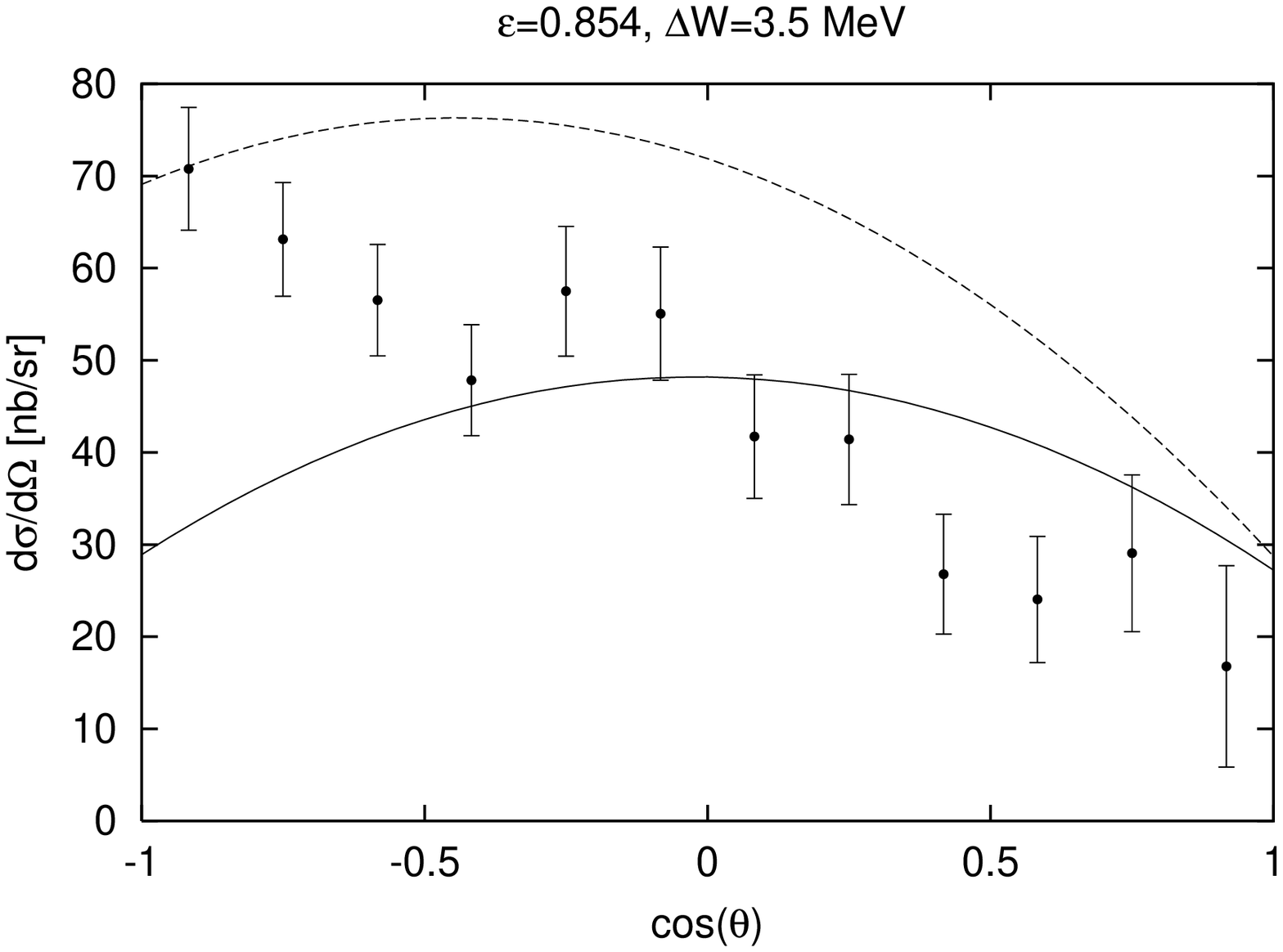,width=2.2in,angle=270}

\vspace{1cm}

\centerline{
\parbox{16cm}{\caption{Differential cross section at $\Delta W =
    2.5\,$MeV (left column) and $\Delta W = 3.5\,$MeV (right
    column) at three different values of the photon polarization
    for the NNLO*-600 wave function in comparison to the 
    MAMI data \protect\cite{Ewald}. Fit~1~(2): dashed (solid) lines.
    \label{fig:dXS2535}}}
}
\end{figure}
\noindent
In Figs.~\ref{fig:dXS0515},\ref{fig:dXS2535} we show the differential cross sections
for fits~1 and 2 employing the NNLO*-600 wave functions in comparison to the MAMI 
data \cite{Ewald}. These two lines corresponding to the two fit procedures can be
considered as a measure of the theoretical uncertainty at this order.
This uncertainty is  comparable to the the experimental errors.
The bell--shape behaviour of the differential cross
sections at the higher values of the pion excess energy is similar to what is 
found in neutral pion photoproduction off protons  and can be traced
back to the large and delta--dominated third order P--wave  LECs $b_{p,n}$,
which are well described in terms of resonance saturation. Within large fluctuations,
the data show more of a backward--forward angle asymmetry. This feature might
be better described when the P-waves have also been worked out to forth order,
but it is fair to state that we do not find sizeable discrepancies between the
data and the theoretical predictions. As in pion
production off the proton, the S--wave multipoles are only dominant
very close to threshold and the P--waves start to dominate at excess
energies of a few MeV.

\medskip\noindent
\begin{figure}[htpb]
   \vspace{0.5cm}
   \centerline{\epsfig{file=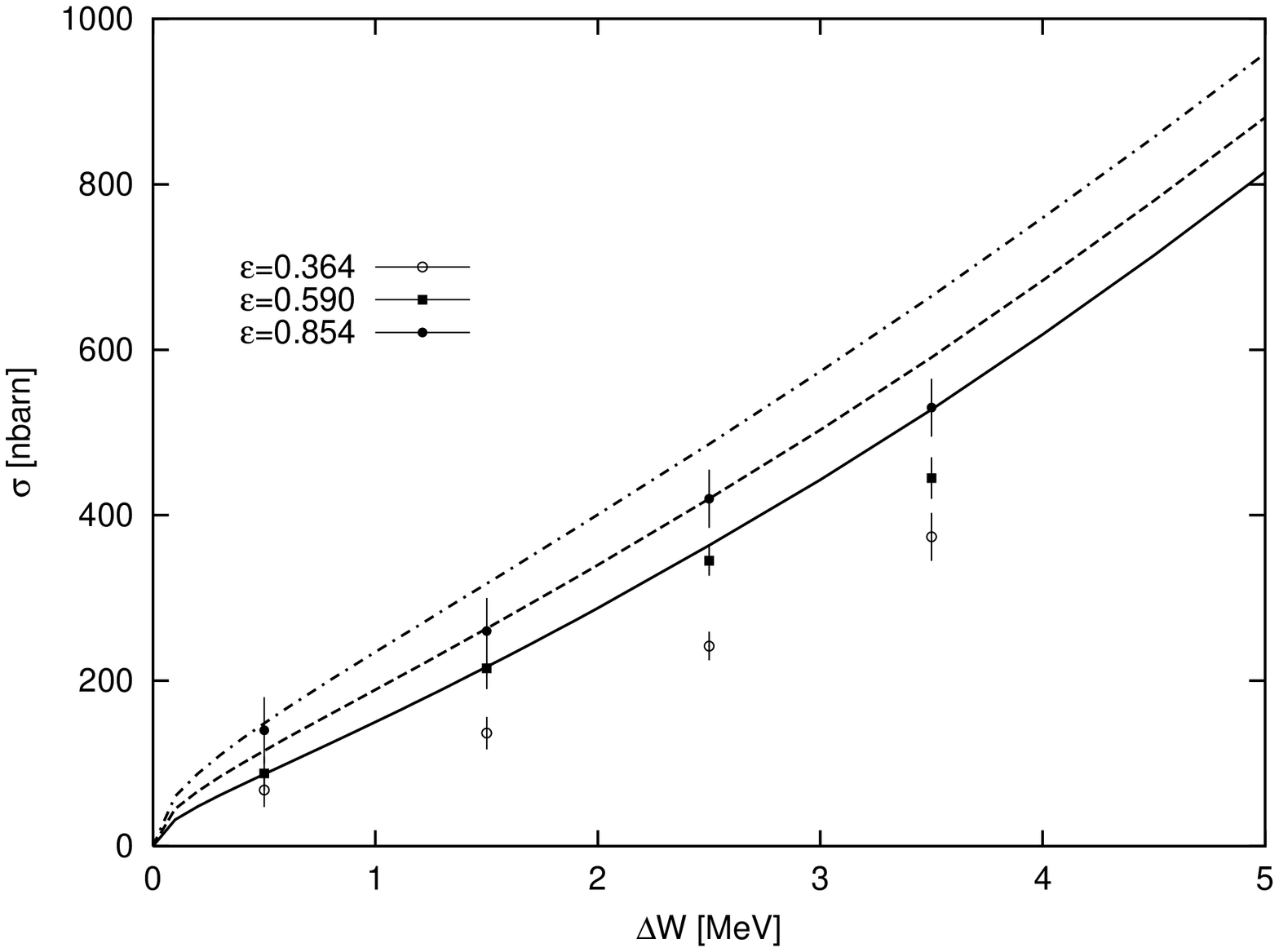,height=4.5in,angle=270}}
   \vspace{0.3cm}
   \centerline{\parbox{11cm}{\caption{\label{fig:tot1}
   Total cross section as a function of $\Delta W$ for three
   different values of the photon polarization in comparison
   to the MAMI data \protect\cite{Ewald} for fit~1 and the
   NNLO*-600 wave function.
  }}}

\vspace{0.5cm}

   \vspace{0.5cm}
   \centerline{\epsfig{file=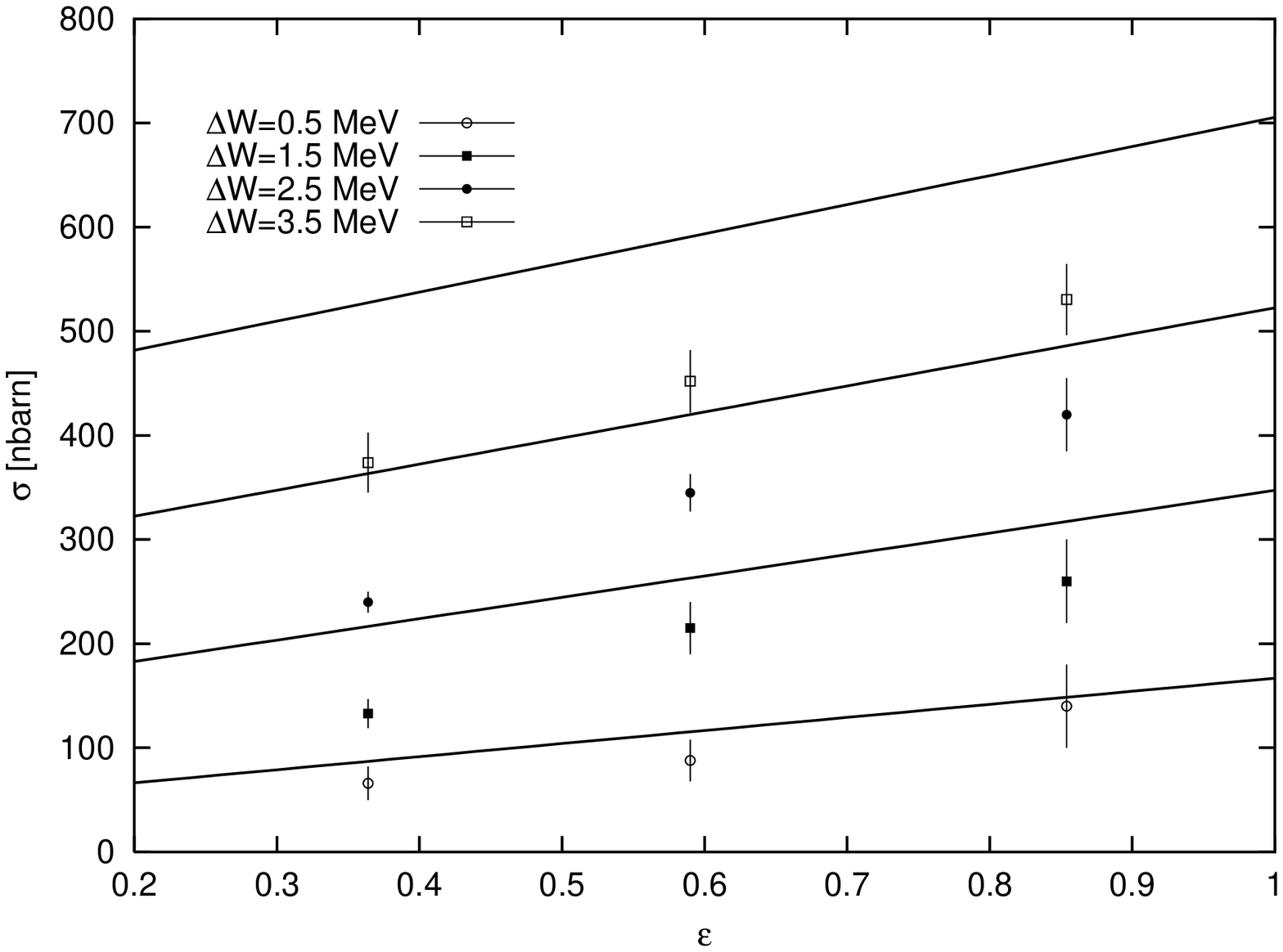,height=4.5in,angle=270}}
   \vspace{0.3cm}
   \centerline{\parbox{11cm}{\caption{\label{fig:ros1}
   Total cross section as a function of the photon polarization 
   $\varepsilon$ for four different values of the pion excess
   energy $\Delta W$ in comparison to the MAMI data
   \protect\cite{Ewald} for fit~1 and the
   NNLO*-600 wave function.
  }}}
\end{figure}
\begin{figure}[htbp]
   \vspace{0.5cm}
   \centerline{\epsfig{file=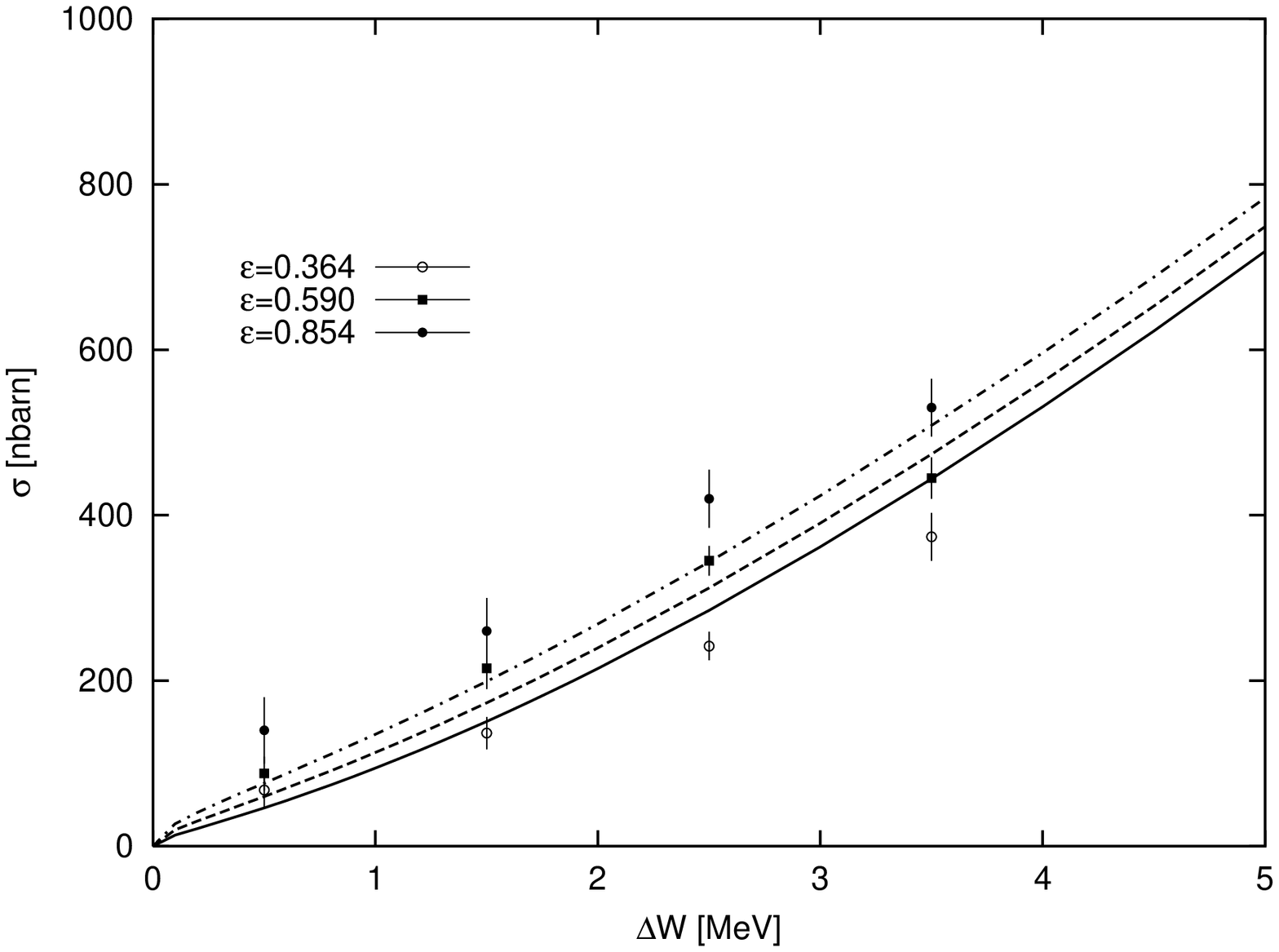,height=4.5in,angle=270}}
   \vspace{0.3cm}
   \centerline{\parbox{11cm}{\caption{\label{fig:tot2}
   Total cross section as a function of $\Delta W$ for three
   different values of the photon polarization in comparison
   to the MAMI data \protect\cite{Ewald} for fit~2 and the
   NNLO*-600 wave function.
  }}}

\vspace{0.5cm}

   \vspace{0.5cm}
   \centerline{\epsfig{file=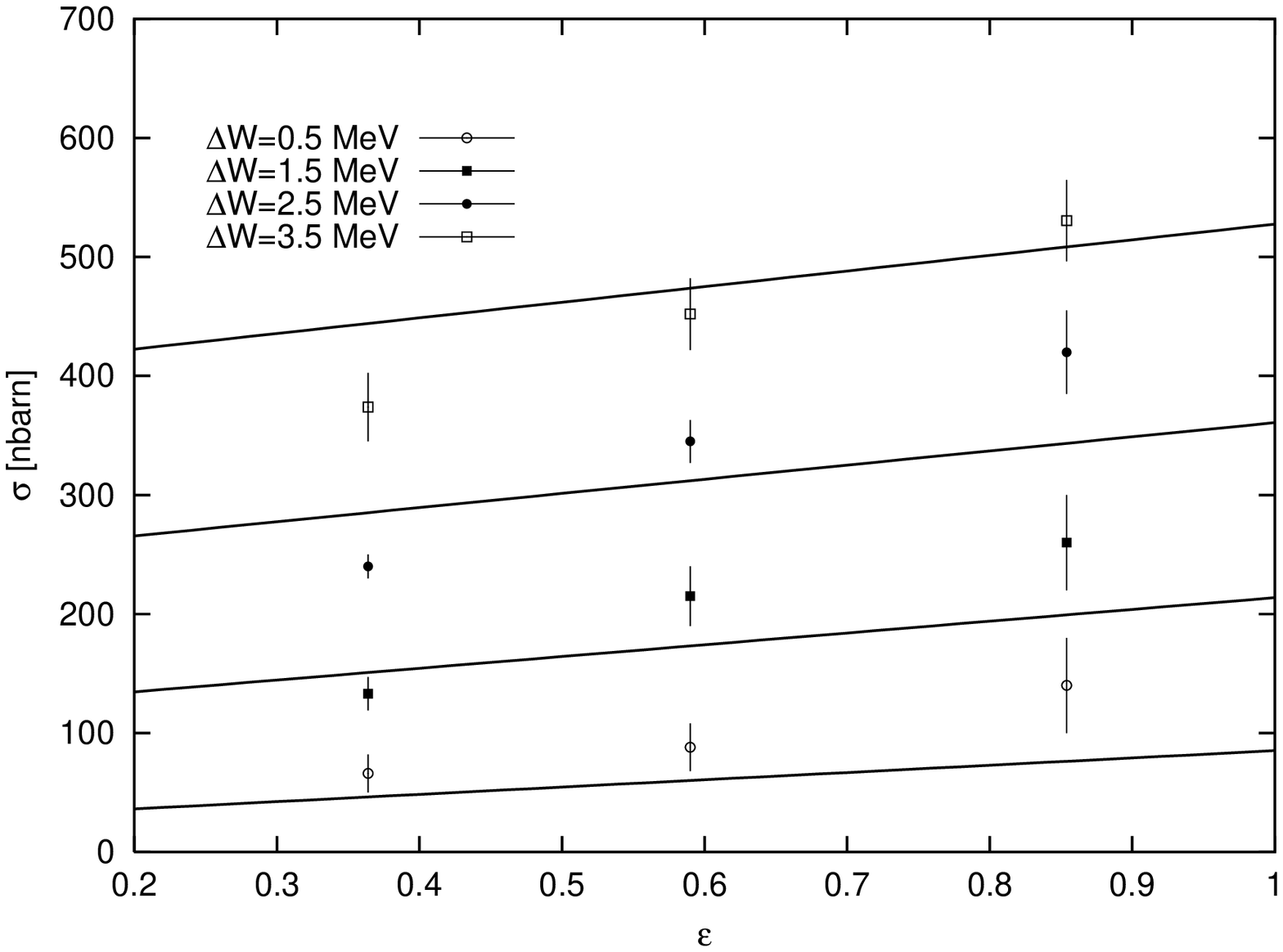,height=4.5in,angle=270}}
   \vspace{0.3cm}
   \centerline{\parbox{11cm}{\caption{\label{fig:ros2}
   Total cross section as a function of the photon polarization 
   $\varepsilon$ for four different values of the pion excess
   energy $\Delta W$ in comparison to the MAMI data
   \protect\cite{Ewald} for fit~2 and the
   NNLO*-600  wave function.
  }}}
\end{figure}
\begin{figure}[ht]
   \vspace{0.5cm}
   \centerline{\epsfig{file=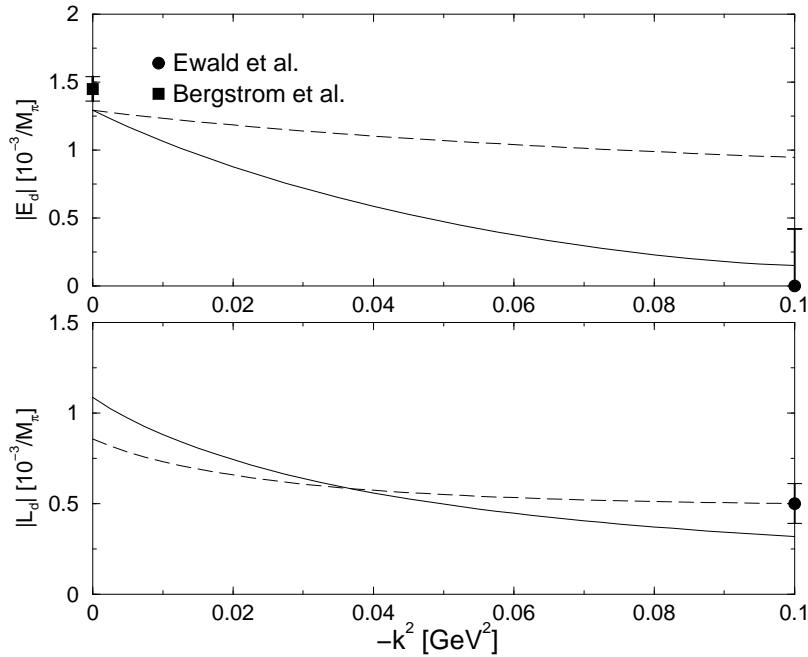,height=3.5in,angle=0}}
   \vspace{0.7cm}
   \centerline{\parbox{11cm}{\caption{\label{fig:ELd}
   Threshold multipoles $|E_d|$ and $|L_d|$ as a function of the
   photon virtuality in comparison to the photon point data
   from SAL \protect\cite{SALd} and the electroproduction data from
   MAMI  \protect\cite{Ewald}. The sign of the experimental result
   for $L_d$ is taken to agree with the theoretical prediction.
   In our fits, the positive sign for $L_d$ is preferred.
   Solid (dashed) lines: Fit~2 (1). The NNLO*-600 wave function is used.
  }}}
\end{figure}

The corresponding total
cross sections as a function of the excess energy $\Delta W$ and of the photon
polarization $\varepsilon$ are shown in Figs.~\ref{fig:tot1} and \ref{fig:ros1}
for fit~1 and in Figs.~\ref{fig:tot2} and \ref{fig:ros2} for fit~2
and the NNLO*-600 wave function, respectively.
We notice that for fit~1 with increasing excess energy and, in particular, with
increasing photon polarization the data are systematically below the chiral 
prediction. Due to the fitting procedure, the slopes of the various curves
for the Rosenbluth separation shown in  Figs.~\ref{fig:ros1} are of course
correct. 
On the other hand, the fitting procedure~2 gives an overall
better description of the total cross sections, cf. Fig.~\ref{fig:tot2} 
with a somewhat too small longitudinal S--wave contribution, as
most clearly seen in Fig.~\ref{fig:ros2}, where again the Rosenbluth separation of
the total cross section is plotted.  These observation can further be
sharpened by considering the dominant longitudinal multipoles as
visualized in Fig.~\ref{fig:ELd}, where the transverse and longitudinal
threshold S--wave multipoles $E_d$ and $L_d$ are shown in comparison to the
data for fits~1 and 2 using the NNLO*-600 wave function.
We also note that $E_d$ is slightly below the data from SAL \cite{SALd},
whereas the prediction from \cite{BBLMvK} was by the same amount above the
data. This can be traced back to a variety of effects. First, we use
slightly different input parameters (for the neutron) so that
the single scattering contribution  is somewhat reduced. Further, in
contrast to Ref.\cite{BBLMvK} we include the pion mass difference in the
three-body contribution, which further reduces $E_d$ by about 
$0.3\cdot 10^{-3}/M_\pi$, compare Fig.~\ref{fig:S} (the energy dependence
of $E_{01}^1$ is almost the same at the photon point $k^2 = 0$). 
Also, our treatment  of the Fermi motion (boost
correction) is improved as compared to that paper and we thus have an
additional reduction of $E_d$. Given that there are other fourth order
effects, our result for the transverse threshold multipole is consistent
with the data. In Table~\ref{tab:a0} we collect the S--wave cross section
$a_{0d}$ for the various wave functions and fit procedures. We remark again
that the scaled S--wave cross section given in \cite{BKM} was much too large
because the dominant longitudinal multipole was not correctly represented.
Thus, the dramatic difference between the CHPT prediction and the data has 
disappeared, and the overall description of the data is satisfactory but
still needs to be improved. This will presumably
be achieved when a complete fourth order calculation including the
dominant isospin breaking effects has been performed.

\renewcommand{\arraystretch}{1.4}
\begin{table}[htb]
\begin{center}
\begin{tabular}{|l||c|c|c|c|c|c|c}
\hline
$-k^2$ [GeV$^2$]      & 0.00 & 0.02 & 0.04 & 0.06 & 0.08 & 0.10 \\
\hline
$a_{0d}$ (NLO-500) [1]   & 0.029 & 0.031 & 0.032 & 0.035 & 0.039 & 0.046 \\ 
$a_{0d}$ (NLO-500) [2]   & 0.029 & 0.022 & 0.016 & 0.014 & 0.013 & 0.013 \\ 
$a_{0d}$ (NLO-600) [1]   & 0.027 & 0.029 & 0.030 & 0.032 & 0.037 & 0.043 \\ 
$a_{0d}$ (NLO-600) [2]   & 0.027 & 0.020 & 0.016 & 0.013 & 0.013 & 0.014 \\ 
$a_{0d}$ (NNLO*-500) [1] & 0.033 & 0.035 & 0.036 & 0.039 & 0.043 & 0.048 \\ 
$a_{0d}$ (NNLO*-500) [2] & 0.033 & 0.024 & 0.018 & 0.014 & 0.013 & 0.012 \\ 
$a_{0d}$ (NNLO*-600) [1] & 0.033 & 0.035 & 0.037 & 0.039 & 0.043 & 0.048 \\ 
$a_{0d}$ (NNLO*-600) [2] & 0.033 & 0.025 & 0.018 & 0.015 & 0.013 & 0.013 \\ 
\hline\end{tabular}
\parbox{11cm}{
\caption{S--wave cross section $a_{0d}$ in $\mu$b for the various wave functions
(w.f.) and fit procedures [$n$] ($n=1,2$ ) employed.
\label{tab:a0}}}
\end{center}
\end{table}

\section{Summary and conclusions}
\def\theequation{\arabic{section}.\arabic{equation}}
\setcounter{equation}{0}
\label{sec:summ}

In this paper, we have studied neutral pion electroproduction off deuterium
in the framework of chiral perturbation theory at and above threshold.
The salient ingredients and results of this work can  be summarized as
follows:
\begin{itemize}
\item[i)]We have developed a general multipole decomposition for
neutral pion production off spin-1 particles that is particularly
suited for the threshold region and  formulated in close analogy
to the standard CGLN amplitudes for pion production off nucleons (spin-1/2 particles).
Similar work was previously published in \cite{Aren1,Aren2}.
\item[ii)]The interaction kernel and the wave functions are
based consistently on chiral effective field theory. The
kernel decomposes into a single scattering and a three--body
contribution. We have chirally expanded the various contributions
working to first non--trivial loop order ${\cal O}(q^3)$, with the
exception of the S--waves for the single scattering contribution.
These have to be included to fourth order with one additional fifth
order term \cite{BKMe}. 
\item[iii)]
All parameters for pion production off the proton
and the ones appearing in the three-body terms are fixed.
The longitudinal neutron S--wave amplitude
contains effectively two parameters, which we have determined
by two different procedures. In the  fits of type~1 
we have fitted the fifth order parameter
to the threshold multipole $L_d$ from Ref.\cite{Ewald}
(and assuming resonance saturation to pin down the other
LEC). The second procedure is based on a two parameter fit 
to the total cross section data from
Ref.\cite{Ewald}. All results are completely insensitive to the
wave functions used, showing that this reaction is sensitive to the
long--range pion exchange firmly rooted in the chiral symmetry of QCD.
\item[iv)] The predicted differential cross sections are satisfactorily
described for both fit procedures, although some systematic discrepancies for
the higher values of the excess energy $\Delta W$ remain, 
see Figs.~\ref{fig:dXS0515},\ref{fig:dXS2535}. In particular, for
fit~1 the total cross section rises too steeply with pion excess energy.
\item[v)]The calculated S-- and P--wave multipoles exhibit a more
complex pion energy and photon virtuality dependence as assumed in the
fits of Ref.~\cite{Ewald}. Within one standard deviation, the chiral predictions
for the threshold multipoles $|E_d|$ and $|L_d|$ are consistent with the
data at $k^2 =0$ \cite{SALd} and $k^2 = -0.1\,$GeV$^2$ \cite{Ewald}. 
\end{itemize}
\medskip\noindent
Clearly, the calculation presented here needs to be improved, in particular,
the fourth order corrections to the P--waves and the three-body terms
have to be included (note that similar work for the P--waves in neutral
pion production off protons has only appeared recently \cite{BKMa}).
However, we have demonstrated that chiral perturbation theory can be used
successfully to analyze pion electroproduction data off the deuteron which gives access
to the elementary neutron amplitude. It would be very interesting to
also have data at lower photon virtuality, which might also help to resolve
the mystery surrouding the proton data at $k^2 = -0.05\,$GeV$^2$.

\section*{Acknowledgements}
We are grateful to Christoph Hanhart for some useful comments.


\bigskip\bigskip

\appendix
\def\theequation{\Alph{section}.\arabic{equation}}
\setcounter{equation}{0}
\section{Equivalence of multipole expansions}
\label{app:equiv}
 
In this appendix, we show the formal equivalence between the
multipole expansion employed here and the earlier one
developed in \cite{Aren2}. Our transition amplitude has the
form (we suppress here the phase factor related to the $\varphi$
dependence)
\beq\label{multus}
t_{m',\lambda,m} (\theta) = \sum_{L_\pi,L,J} 
\langle L_\pi \, \lambda+m-m' \, 1 \, m' \, | \, J \,
\lambda+m \, \rangle \langle 1 \, m \, L \, \lambda
\, | \, J \, \lambda+m \, \rangle \, {\cal O}_{L_\pi,J}^{L,\lambda}
\, Y_{L_\pi,\lambda+m-m'} (\theta, \varphi=0)
\eeq
in terms of the spherical harmonics. Arenh\"ovel \cite{Aren2} works in the 
helicity basis and uses the rotation matrices $d^J_{M,M'} (\theta)$,
\beq\label{multA}
t_{m',\lambda,m}^A (\theta) = \sum_{L_\pi,L,J} 
\langle L_\pi \, 0 \, 1 \, -m' \, | \, J \,
-m' \, \rangle \langle 1 \, -m \, L \, \lambda
\, | \, J \, \lambda-m \, \rangle \, \frac{\hat{L}_\pi}{\sqrt{4\pi}}\,
{\cal O}_{L_\pi,J}^{L,\lambda}
\, d_{\lambda-m,-m'}^J (\theta)~.
\eeq
We now show that
\beq
t_{m',\lambda,m}^A (\theta) = \sum_{\mu=-1}^1 \, 
t_{-\mu,\lambda,-m} (\theta)\, d_{-\mu,-m'}^1 (\theta)~.
\eeq
This can be proven simply by using the relation between the
spherical harmonics and the d--functions,
\beq
Y_{L_\pi,\lambda+m-m'} (\theta, \varphi=0) = 
\frac{\hat{L}_\pi}{\sqrt{4\pi}}\, d^{L_\pi}_{\lambda+m-m',0} (\theta)~,
\eeq
and the following relation between rotation matrices
\beq
\sum_{m_1,m_2} \langle j_1 \, m_1\, j_2 \, m_2\, |\, j\,m \,\rangle
\, d^{j_1}_{m_1,m_1'} (\theta) \, d^{j_2}_{m_2,m_2'} (\theta) =
\sum_{m'} \langle j_1 \, m_1'\, j_2 \, m_2'\, |\, j\,m' \,\rangle 
\, d^{j}_{m,m'} (\theta)~.
\eeq
With this, the equivalence between Eqs.~(\ref{multus}) and
(\ref{multA}) follows immediately.
 
\setcounter{equation}{0}
\section{Transformation matrices}
\label{app:mult}

In this appendix, we collect the expressions of the various matrices
appearing in Eqs.~(\ref{matDE},\ref{matGH}). For notational simplicity,
we substitute the symbol for the pion angular momentum $L_\pi$ by $L$ in 
this appendix. Consider first $D_L$. The non--zero matrix elements are:
\beqa
 D_{12} &=&\frac{\frac{i }{8}\,\left( -1 + L \right)
      \,\left( {P_{-2 + L}} - {P_L} \right) }
   {{\sqrt{L}}\,\left( -1 + 2\,L \right) } ~, \quad
 D_{13}  = \frac{\frac{-i }{8}\,\left( -1 + L \right) \,{P_L}}
   {{\sqrt{L}}} ~, \quad 
 D_{14} = \frac{\frac{-i }{8}\,\left( -1 + L \right) \,{P_{-1
   + L}}}{{\sqrt{L}}}~, \quad 
\nonumber \\
 D_{15} &=& \frac
     {-\left( \left( 1 + L \right) \,\left( {P_{-1 + L}} 
     - {P_{1 + L}} \right)  \right) }{8\,{\sqrt{L}}\,
     \left( 1 + 2\,L \right) }~, \quad
 D_{16} =\frac{-\left( \left( 1 + L \right) \,
       \left( {P_{-2 + L}} - {P_L} \right)  \right) }{16\,{\sqrt{L}}\,
      \left( -1 + 2\,L \right) }~, \quad 
 D_{18} = \frac{\left( 1 + L \right) \,{P_L}}{16\,{\sqrt{L}}}~,
    \quad 
\nonumber \\
  D_{19} &=& \frac{\left( 1 + L \right) \,{P_{-1 + L}}}
   {16\,{\sqrt{L}}}~, \quad 
\nonumber \\
 D_{22} &=& \frac{-i \,{\sqrt{\frac{-1 + L}{L}}}\,{\sqrt{1 + L}}\,
     \left( {P_{-2 + L}} - {P_L} \right) }{-8 + 16\,L}~,\quad
 D_{23} =  \frac{i }{8}\,{\sqrt{\frac{-1 + L}{L}}}\,
   {\sqrt{1 + L}}\,{P_L}  ~, \quad
 D_{24} = \frac{i }{8}\,{\sqrt{\frac{-1 + L}{L}}}\,{\sqrt{1 + L}}\,
   {P_{-1 + L}}~, \quad 
\nonumber \\
 D_{25} &=&  \frac{{\sqrt{\frac{-1 + L}{L}}}\,{\sqrt{1 + L}}\,
     \left( -{P_{-1 + L}} + {P_{1 + L}} \right) }{8 + 16\,L}~, \quad 
 D_{26} =  \frac{{\sqrt{\frac{-1 + L}{L}}}\,
     {\sqrt{1 + L}}\,\left( -{P_{-2 + L}} + {P_L} \right) }{-16 +
       32\,L}~,  
\nonumber \\ 
  D_{28} &=& \frac{
     {\sqrt{\frac{-1 + L}{L}}}\,{\sqrt{1 + L}}\,{P_L}}{16}~, \quad
  D_{29} =  \frac{{\sqrt{\frac{-1 + L}{L}}}\,
     {\sqrt{1 + L}}\,{P_{-1 + L}}}{16}~,  
\nonumber \\
   D_{32} &=&\frac{i \,{\sqrt{\frac{L}{1 + L}}}\,{\sqrt{2 + L}}\,
     \left( {P_L} - {P_{2 + L}} \right) }{24 + 16\,L}~, \quad
 D_{33} =  \frac{-i }{8}\,{\sqrt{\frac{L}{1 + L}}}\,
   {\sqrt{2 + L}}\,{P_L}~, \quad
 D_{34} = \frac{-i }{8}\,{\sqrt{\frac{L}{1 + L}}}\,{\sqrt{2 + L}}\,
   {P_{1 + L}} ~, 
\nonumber\\
 D_{35} &=& \frac{{\sqrt{\frac{L}{1 + L}}}\,{\sqrt{2 + L}}\,
     \left( {P_{-1 + L}} - {P_{1 + L}} \right) }{8 + 16\,L}~, \quad
 D_{36} =  \frac{{\sqrt{\frac{L}{1 + L}}}\,
     {\sqrt{2 + L}}\,\left( {P_L} - {P_{2 + L}} \right) }{48 + 32\,L}~, 
\nonumber\\
  D_{38} &=& \frac{-\left( 
       {\sqrt{\frac{L}{1 + L}}}\,{\sqrt{2 + L}}\,{P_L} \right) }{16}~, \quad
  D_{39} =  \frac{-\left( 
       {\sqrt{\frac{L}{1 + L}}}\,{\sqrt{2 + L}}\,{P_{1 + L}} \right)
   }{16}~, 
\nonumber \\ 
 D_{42} &=&  \frac{ \frac{-i }{8}\,\left( 2 + L \right) \,\left( {P_L} 
      - {P_{2 + L}} \right) }{{\sqrt{1 + L}}\, \left( 3 + 2\,L \right) }~,
     \quad
 D_{43} =  \frac{\frac{i }{8}\,\left( 2 + L \right) \,{P_L}}
   {{\sqrt{1 + L}}}~, \quad
 D_{44} = \frac{\frac{i }{8}\,\left( 2 + L \right) \,{P_{1 + L}}}{{\sqrt{1
   + L}}} ~, 
\nonumber\\
 D_{45} &=& \frac{L\,\left( {P_{-1 + L}} - {P_{1 + L}} \right) }
       {8\,{\sqrt{1 + L}}\,\left( 1 + 2\,L \right) }~, \quad
 D_{46} = \frac{L\,\left( {P_L} - {P_{2 + L}} \right) }
   {16\,{\sqrt{1 + L}}\,\left( 3 + 2\,L \right) } ~,  \quad
 D_{48}  = \frac{-\left( L\,{P_L} \right) }
   {16\,{\sqrt{1 + L}}}~, \quad
  D_{49} =   \frac{-\left( L\,{P_{1 + L}} \right) }
   {16\,{\sqrt{1 + L}}}~, 
\nonumber \\
 D_{55} &=& \frac{{\sqrt{-2 + L}}\,{\sqrt{L}}\,\left( 1 + L \right) \,
     \left( {P_{-1 + L}} - {P_{1 + L}} \right) }{8\,{\sqrt{-1 +
   2\,L}}\,\left( 1 + 2\,L \right) }~, \quad
 D_{56} =  
    \frac{{\sqrt{-2 + L}}\,L^{\frac{3}{2}}\,\left( {P_{-2 + L}} - {P_L} \right) }
   {8\,{\left( -1 + 2\,L \right) }^{\frac{3}{2}}} ~,  
\nonumber \\
 D_{57} &=& \frac{{\sqrt{-2 + L}}\,\left( -1 + L \right) \,
     {\sqrt{L}}\,\left( {P_{-3 + L}} - {P_{-1 + L}} \right) }
     {4\,\left( -3 + 2\,L \right) \,{\sqrt{-4 + 8\,L}}}~, \quad
  D_{58} = \frac{-\left( {\sqrt{-2 + L}}\,{\sqrt{L}}\,{P_L} \right) }
   {8\,{\sqrt{-1 + 2\,L}}}~, 
\nonumber\\ 
  D_{59} &=&  \frac{-\left( {\sqrt{-2 + L}}\,{\sqrt{L}}\,{P_{-1 + L}} \right) }
   {8\,{\sqrt{-1 + 2\,L}}}~, 
\nonumber\\ 
  D_{61} &=& \frac{{\sqrt{L}}\,{\sqrt{1 + L}}\,{\sqrt{-1 + 2\,L}}\,
     \left( {P_{-1 + L}} - {P_{1 + L}} \right) }{8 + 16\,L}~, 
\nonumber\\ 
  D_{62} &=& \frac{\frac{i }{8}\,
     \left( \left( -3 - 2\,L + 3\,L^2 + 2\,L^3 \right) \,{P_{-2 + L}} + 
       \left( 3 + 4\,L - 6\,L^2 - 4\,L^3 \right) \,{P_L} + 
       L\,\left( -2 + 3\,L + 2\,L^2 \right) \,{P_{2 + L}} \right)
    }{{\sqrt{\frac{L}{1 + L}}}\,
     {\sqrt{-1 + 2\,L}}\,\left( 1 + 2\,L \right) \,\left( 3 + 2\,L
      \right) }~, 
\nonumber \\
 D_{63} &=& \frac{\frac{i }{8}\,
     {\sqrt{-1 + 2\,L}}\,{P_L}}{{\sqrt{\frac{L}{1 + L}}}}~, \quad
 D_{64} = \frac{\frac{i }{8}\,{\sqrt{-1 + 2\,L}}\,
     \left( \left( 1 + L \right) \,{P_{-1 + L}} + L\,{P_{1 + L}}
   \right) }{{\sqrt{\frac{L}{1 + L}}}\,  \left( 1 + 2\,L \right) }~,
\nonumber \\
 D_{65} &=& \frac{-\left( \left( -3 + L + L^2 \right) \,
       \left( {P_{-1 + L}} - {P_{1 + L}} \right)  \right) }{24\,{\sqrt{\frac{L}{1 + L}}}\,
     {\sqrt{-1 + 2\,L}}\,\left( 1 + 2\,L \right) }~, 
\nonumber \\
 D_{66} &=&  \frac{-\left( {\left( 3 + 2\,L \right) }^2\,
        \left( -1 + L^2 \right) \,{P_{-2 + L}} \right)  + 
     \left( -9 - 14\,L + 12\,L^2 + 8\,L^3 \right) \,{P_L} + 
     {\left( 1 - 2\,L \right) }^2\,L\,\left( 2 + L \right) \,{P_{2 +
    L}}}{48\,{\sqrt{\frac{L}{1 + L}}}\,
     {\left( -1 + 2\,L \right) }^{\frac{3}{2}}\,\left( 1 + 2\,L
       \right) \,\left( 3 + 2\,L \right) }~,  
\nonumber \\  
 D_{67} &=&
   \frac{{\sqrt{L}}\,{\left( 1 + L \right) }^{\frac{3}{2}}\,
        \left( -{P_{-1 + L}} + {P_{1 + L}} \right) }
   {24\,{\sqrt{-1 + 2\,L}}\,\left( 1 + 2\,L \right) }~,  \quad
  D_{68} =  \frac{-{P_L}}
   {16\,{\sqrt{\frac{L}{1 + L}}}\,{\sqrt{-1 + 2\,L}}}~, 
\nonumber \\
  D_{69} &=&  \frac{-\left( \left( 3 + 5\,L + 2\,L^2 \right) \,
        {P_{-1 + L}} + \left( 1 - 2\,L \right) \,L\,{P_{1 + L}}
    \right) } {48\,{\sqrt{\frac{L}{1 + L}}}\,
     {\sqrt{-1 + 2\,L}}\,\left( 1 + 2\,L \right) }~, 
\nonumber\\  
    D_{71} &=&\frac{{\sqrt{L}}\,{\sqrt{1 + L}}\,
     \left( -{P_{-1 + L}} + {P_{1 + L}} \right) }{8\,{\sqrt{1 +
      2\,L}}}~, 
\nonumber \\ 
    D_{72} &=& \frac{\frac{-i }{8}\,
     \left( \left( -3 - 2\,L + 3\,L^2 + 2\,L^3 \right) \,{P_{-2 + L}} + 
       \left( 3 + 4\,L - 6\,L^2 - 4\,L^3 \right) \,{P_L} + 
       L\,\left( -2 + 3\,L + 2\,L^2 \right) \,{P_{2 + L}} \right) }
     {{\sqrt{L}}\,{\sqrt{1 + L}}\,
     {\sqrt{1 + 2\,L}}\,\left( -3 + 4\,L + 4\,L^2 \right) }~,
\nonumber \\
 D_{73} &=&   \frac{\frac{-i }{8}\,
     {\sqrt{1 + 2\,L}}\,{P_L}}{{\sqrt{L}}\,{\sqrt{1 + L}}}~, \quad
 D_{74} =  \frac{\frac{-i }{8}\,
     \left( \left( 1 + L \right) \,{P_{-1 + L}} + L\,{P_{1 + L}}
   \right) } {{\sqrt{L}}\,{\sqrt{1 + L}}\,
     {\sqrt{1 + 2\,L}}}  ~, 
\nonumber \\
 D_{75} &=& \frac{-\left( \left( -3 + L + L^2 \right) \,
       \left( {P_{-1 + L}} - {P_{1 + L}} \right)  \right) }{24\,{\sqrt{L}}\,
         {\sqrt{1 + L}}\, {\sqrt{1 + 2\,L}}}~, 
\nonumber \\
 D_{76} &=& \frac{-\left( {\left( 3 + 2\,L \right) }^2\,\left( -1 + L^2 \right) \,
        {P_{-2 + L}} \right)  + \left( -9 - 14\,L + 12\,L^2 + 8\,L^3 \right) \,{P_L} + 
     {\left( 1 - 2\,L \right) }^2\,L\,\left( 2 + L \right) 
      \,{P_{2 + L}}}{48\,{\sqrt{L}}\,
     {\sqrt{1 + L}}\,{\sqrt{1 + 2\,L}}\,\left( -3 + 4\,L + 4\,L^2
       \right) }~,  
\nonumber \\
 D_{77} &=& \frac{{\sqrt{L}}\,
     {\sqrt{1 + L}}\,\left( -{P_{-1 + L}} + {P_{1 + L}} \right)
       }{24\,{\sqrt{1 + 2\,L}}}~, \quad
 D_{78} = \frac{-\left(
       {\sqrt{1 + 2\,L}}\,{P_L} \right) }{16\,{\sqrt{L}}\,{\sqrt{1 +
       L}}}~, 
\nonumber \\
 D_{79} &=&  \frac{-\left( 
        \left( 3 + 5\,L + 2\,L^2 \right) \,{P_{-1 + L}} 
         + \left( 1 - 2\,L \right) \,L\,{P_{1 + L}}
       \right) }{48\,{\sqrt{L}}\,{\sqrt{1 + L}}\,{\sqrt{1 + 2\,L}}}~,
\nonumber\\  
 D_{81} &=& \frac{{\sqrt{L}}\,
     {\sqrt{1 + L}}\,{\sqrt{3 + 2\,L}}\,\left( {P_{-1 + L}} 
      - {P_{1 + L}} \right) }{8 + 16\,L}~, 
\nonumber \\
 D_{82} &=& \frac{\frac{-i }{8}\,{\sqrt{\frac{L}{1 + L}}}\,
   \left( \left( -3 - 2L + 3L^2 + 2L^3 \right) \,{P_{-2 + L}} + 
    \left( 3 + 4L - 6L^2 - 4L^3 \right) \,{P_L} + 
   L\,\left( -2 + 3L + 2L^2 \right) \,{P_{2 + L}} \right) }{{\sqrt{3 + 2L}}\,
   \left( -1 + 4L^2 \right)}~, 
\nonumber\\ 
 D_{83} &= &\frac{-i }{8}\,{\sqrt{\frac{L}{1 + L}}}\,{\sqrt{3 + 2\,L}}\,
   {P_L} ~, \quad
 D_{84} = \frac{-i \,{\sqrt{\frac{L}{1 + L}}}\,{\sqrt{3 + 2\,L}}\,
     \left( \left( 1 + L \right) \,{P_{-1 + L}} + L\,{P_{1 + L}}
   \right) }{8 + 16\,L} ~, 
\nonumber\\ 
 D_{85} &=& \frac{-\left(
       {\sqrt{\frac{L}{1 + L}}}\,\left( -3 + L + L^2 \right) \,
       \left( {P_{-1 + L}} - {P_{1 + L}} \right)  \right) }{24\,\left( 1 
     + 2\,L \right) \,{\sqrt{3 + 2\,L}}}~, 
\nonumber \\
 D_{86} &=&  \frac{-\left( {\sqrt{\frac{L}{1 + L}}}\,
       \left( {\left( 3 + 2\,L \right) }^2\,\left( -1 + L^2 \right) \,{P_{-2 + L}} + 
         \left( 9 + 14\,L - 12\,L^2 - 8\,L^3 \right) \,{P_L} - 
     {\left( 1 - 2\,L \right) }^2\,L\,\left
     ( 2 + L \right) \,{P_{2 + L}} \right)  \right) }{48\,
     {\left( 3 + 2\,L \right) }^{\frac{3}{2}}\,\left( -1 + 4\,L^2
       \right) }~,  
\nonumber \\ 
 D_{87} &=& \frac{L^{\frac{3}{2}}\,
     {\sqrt{1 + L}}\,\left( -{P_{-1 + L}} + {P_{1 + L}} \right) }{24\,\left( 1 
      + 2\,L \right) \,
     {\sqrt{3 + 2\,L}}} ~, \quad
  D_{88} = \frac{-\left( {\sqrt{\frac{L}{1 + L}}}\,{P_L} \right) }
   {16\,{\sqrt{3 + 2\,L}}}~, 
\nonumber \\
 D_{89} &=&  \frac{-\left( {\sqrt{\frac{L}{1 + L}}}\,
       \left( \left( 3 + 5\,L + 2\,L^2 \right) \,{P_{-1 + L}} + 
         \left( 1 - 2\,L \right) \,L\,{P_{1 + L}} \right)  \right) }
       {48\,\left( 1 + 2\,L \right) \,
     {\sqrt{3 + 2\,L}}} ~, 
\nonumber\\
 D_{95} &=& \frac{L\,{\sqrt{1 + L}}\,{\sqrt{3 + L}}\,
     \left( {P_{-1 + L}} - {P_{1 + L}} \right) }{8\,\left( 1 + 2\,L
   \right) \,{\sqrt{3 + 2\,L}}}~,  \quad
 D_{96} =  
    \frac{{\left( 1 + L \right) }^{\frac{3}{2}}\,{\sqrt{3 + L}}\,
     \left( {P_L} - {P_{2 + L}} \right) }
   {8\,{\left( 3 + 2\,L \right) }^{\frac{3}{2}}}~,  
\nonumber \\
 D_{97} &=& \frac{{\sqrt{1 + L}}\,\left( 2 + L \right) \,
     {\sqrt{3 + L}}\,\left( {P_{1 + L}} - {P_{3 + L}} \right) }{8\,{\sqrt{3 + 2\,L}}\,
     \left( 5 + 2\,L \right) }~, \quad
 D_{98} = \frac{{\sqrt{1 + L}}\,{\sqrt{3 + L}}\,{P_L}}
   {8\,{\sqrt{3 + 2\,L}}}~, 
\nonumber \\
 D_{99} &=& \frac{{\sqrt{1 + L}}\,{\sqrt{3 + L}}\,{P_{1 + L}}}
   {8\,{\sqrt{3 + 2\,L}}}~. \nonumber
\eeqa
\beqa 
\eeqa
Here, the $P_L$ are the conventional Legendre polynomials that depend
on $z = \cos \theta$. Note that $L$ is positive definite so that Legendre
polynomials with negative index have to be understood as zero. The matrix
$E_L$ has no zero elements and takes the form
\beqa
E_{L}=\left(\matrix{ \frac{\frac{-i }{4}\,{\sqrt{-1 +
          L}}\,{P_L}}{{\sqrt{2}}} & \frac{\frac{-i }{4}\,
     {\sqrt{-1 + L}}\,{P_{-1 + L}}}{{\sqrt{2}}} & \frac{{\sqrt{-1 +
         L}}\,\left( 1 + L \right) \,
     \left( {P_{-1 + L}} - {P_{1 + L}} \right) }{8\,{\sqrt{2}}\,\left(
       1 + 2\,L \right) } & \frac{
     {\sqrt{-1 + L}}\,\left( 1 + L \right) \,\left( {P_{-2 + L}} -
      {P_L} \right) }{8\,{\sqrt{2}}\,
     \left( -1 + 2\,L \right) } \cr \frac{\frac{i }{4}\,{\sqrt{1 +
         L}}\,{P_L}}{{\sqrt{2}}} & 
    \frac{\frac{i }{4}\,{\sqrt{1 + L}}\,{P_{-1 + L}}}{{\sqrt{2}}}
    & \frac{\left( -1 + L \right) \,
     {\sqrt{1 + L}}\,\left( {P_{-1 + L}} - {P_{1 + L}} \right)
     }{8\,{\sqrt{2}}\,\left( 1 + 2\,L \right) }
   & \frac{\left( -1 + L \right) \,{\sqrt{1 + L}}\,\left( {P_{-2 + L}}
       - {P_L} \right) }
   {8\,{\sqrt{2}}\,\left( -1 + 2\,L \right) } \cr \frac{\frac{i
       }{4} \,{\sqrt{L}}\,{P_L}}
   {{\sqrt{2}}} & \frac{\frac{i }{4}\,{\sqrt{L}}\,{P_{1 +
         L}}}{{\sqrt{2}}} & \frac{-\left( 
       {\sqrt{L}}\,\left( 2 + L \right) \,\left( {P_{-1 + L}} - {P_{1
             + L}} \right)  \right) }{8\,
     {\sqrt{2}}\,\left( 1 + 2\,L \right) } & \frac{-\left(
       {\sqrt{L}}\,\left( 2 + L \right) \,
       \left( {P_L} - {P_{2 + L}} \right)  \right) }{8\,{\sqrt{2}}\,
    \left( 3 + 2\,L \right) } \cr 
    \frac{\frac{-i }{4}\,{\sqrt{2 + L}}\,{P_L}}{{\sqrt{2}}} 
    & \frac{\frac{-i }{4}\,
     {\sqrt{2 + L}}\,{P_{1 + L}}}{{\sqrt{2}}} & 
                  \frac{-\left( L\,{\sqrt{2 + L}}\,
       \left( {P_{-1 + L}} - {P_{1 + L}} \right)  \right) }
              {8\,{\sqrt{2}}\,\left( 1 + 2\,L \right) } &
   \frac{-\left( L\,{\sqrt{2 + L}}\,\left( {P_L} - {P_{2 + L}} 
    \right)  \right) }
   {8\,{\sqrt{2}}\,\left( 3 + 2\,L \right) } \cr  }\right)~.
\eeqa

\medskip\noindent
We now turn to the matrices $G_{L_{\pi}}$ and  $H_{L_{\pi}}$ appearing
in the inverse transformation, Eq.~(\ref{matGH}). It is most convenient
to express these with the help of Clebsch--Gordan coefficients. For that,
we employ the $D$--symbols
\begin{eqnarray}
D^{L,L_{\pi},J}_{\lambda,m^{\prime},m}&=& \langle L_{\pi}\,\lambda+m-m^{\prime}\,
1\, m^{\prime}|J\, \lambda+m \rangle \langle 1\,m\,L\,\lambda|J\,\lambda+m
\rangle~,
\end{eqnarray}
as they appear also in the multipole expansion of the $T$--matrix. We define
the transverse and the longitudinal vectors $T(L,L_{\pi},J)$ and
$L(L,L_{\pi},J)$, respectively, in terms of their components:
\begin{eqnarray}
T(L,L_{\pi},J)_{1}&=&\frac{4 \sqrt{1+2 L} \sqrt{1+2 L_{\pi}} 
\left(D^{L,L_{\pi},J}_{1,-1,-1} 
+ D^{L,L_{\pi},J}_{1,0,0} + D^{L,L_{\pi},J}_{1,1,1}\right)
P^{(1)}_{L_{\pi}}}{3 \sqrt{1+2 J}
 \sqrt{L_{\pi} \left(1+L_{\pi}\right)}}~, \nonumber\\
T(L,L_{\pi},J)_{2}&=&\frac{2 i \sqrt{2+4 L} \sqrt{1+2 L_{\pi}}
\left( D^{L,L_{\pi},J}_{1,-1,0} + D^{L,L_{\pi},J}_{1,0,1}\right) 
P^{(2)}_{L_{\pi}}}{\sqrt{1+2 J}\sqrt{L_{\pi}\left(1+L_{\pi}\right)}
\sqrt{-2+L_{\pi}+L_{\pi}^2}}~,\nonumber\\
T(L,L_{\pi},J)_{3}&=&\frac{i \sqrt{2+4 L}\sqrt{1+2 L_{\pi}}(D^{L,L_{\pi},J}_{1,0,-1}
+D^{L,L_{\pi},J}_{1,1,0}) P_{L_{\pi}}}{\sqrt{1+2 J}}\nonumber  \\
&+&\frac{2 i {\sqrt{1+2 L}}\sqrt{1+2 L_{\pi}} z (D^{L,L_{\pi},J}_{1,-1,
-1}-D^{L,L_{\pi},J}_{1,1,1})P^{(1)}_{L_{\pi}}}{\sqrt{1+2 J}
\sqrt{L_{\pi}(1+L_{\pi})}}\nonumber\\
&-&\frac{i {\sqrt{2+4 L}}\sqrt{1+2 L_{\pi}} (-1+z^2)(D^{L,L_{\pi},J}_{1,-1,0}
+D^{L,L_{\pi},J}_{1,0,1})P^{(2)}_{L_{\pi}}}{\sqrt{1+2 J}\sqrt{L_{\pi}
(1+L_{\pi})}\sqrt{-2+L_{\pi}+{{L_{\pi}}^2}}}~,\nonumber\\
T(L,L_{\pi},J)_{4}&=&-\frac{2 i \sqrt{1+2 L}\sqrt{1+2 L_{\pi}}
(D^{L,L_{\pi},J}_{1,-1,-1}-D^{L,L_{\pi},J}_{1,1,1})P^{(1)}_{L_{\pi}}}{\sqrt{1+2 J}
\sqrt{L_{\pi}
(1+L_{\pi})}}~,\nonumber\\
T(L,L_{\pi},J)_{5}&=&\frac{2\sqrt{1+2 L}\sqrt{1+2
L_{\pi}}}{\sqrt{1+2
J}\sqrt{L_{\pi}(1+L_{\pi})}}\left(D^{L,L_{\pi},J}_{1,-1,-1}
-2 D^{L,L_{\pi},J}_{1,0,0}-D^{L,L_{\pi},J}_{1,1,-1}+D^{L,L_{\pi},J}_{1,1,1}\right)
 P^{(1)}_{L_{\pi}}\nonumber\\
&+&\frac{4 \sqrt{2+4 L}\sqrt{1+2 L_{\pi}} z\left(-D^{L,L_{\pi},J}_{1,-1,0}
+D^{L,L_{\pi},J}_{1,0,1}\right)P^{(2)}_{L_{\pi}}}{\sqrt{1+2 J}\sqrt{L_{\pi} 
(1+L_{\pi})}\sqrt{-2+L_{\pi}+{{L_{\pi}}^2}}}\nonumber\\
&+&\frac{2\sqrt{1+2 L}\sqrt{1+2 L_{\pi}}(1+3
z^2)D^{L,L_{\pi},J}_{1,-1,1} P^{(3)}_{L_{\pi}}}{\sqrt{1+2 J}\sqrt{L_{\pi} 
(1+L_{\pi})}\sqrt{-6+L_{\pi}+{{L_{\pi}}^2}}\sqrt{-2+L_{\pi}+{{L_{\pi}}^2}}}~,
\nonumber\\
T(L,L_{\pi},J)_{6}&=&\frac{4\sqrt{2+4 L}\sqrt{1+2 L_{\pi}}(D^{L,L_{\pi},J}_{1,-1,0}
-D^{L,L_{\pi},J}_{1,0,1})P^{(2)}_{L_{\pi}}}{\sqrt{1+2 J}\sqrt{L_{\pi}(1+L_{\pi})}
\sqrt{-2+L_{\pi}
+{L_{\pi}}^2}}\nonumber\\
&-&\frac{16\sqrt{1+2 L}\sqrt{1+2 L_{\pi}} z
D^{L,L_{\pi},J}_{1,-1,1}P^{(3)}_{L_{\pi}}}{\sqrt{1+2 J}\sqrt{L_{\pi}(1+L_{\pi})}
\sqrt{-6+L_{\pi}+{{L_{\pi}}^2}}\sqrt{-2+L_{\pi}+{{L_{\pi}}^2}}}~,\nonumber\\
T(L,L_{\pi},J)_{7}&=&\frac{8\sqrt{1+2 L}\sqrt{1+2 L_{\pi}}  D^{L,L_{\pi},J}_{1,-1,1}
 P^{(3)}_{L_{\pi}}}{\sqrt{1+2 J}\sqrt{L_{\pi} (1+L_{\pi})}
\sqrt{-6+L_{\pi}+{{L_{\pi}}^2}}\sqrt{-2+L_{\pi}+{{L_{\pi}}^2}}}~,\nonumber\\
T(L,L_{\pi},J)_{8}&=&\frac{2 \sqrt{2+4 L}\sqrt{1+2
L_{\pi}}(D^{L,L_{\pi},J}_{1,0,-1}-D^{L,L_{\pi},J}_{1,1,0})P_{L_{\pi}}}{\sqrt{1+2 J}}
\nonumber\\
&+&\frac{4\sqrt{1+2 L}\sqrt{1+2 L_{\pi}} z
D^{L,L_{\pi},J}_{1,1,-1} P^{(1)}_{L_{\pi}}}{\sqrt{1+2 J}\sqrt{L_{\pi}
(1+L_{\pi})}}\nonumber\\
&+&\frac{2\sqrt{2+4 L}\sqrt{1+2
L_{\pi}}(-1+z^2)(D^{L,L_{\pi},J}_{1,-1,0}-D^{L,L_{\pi},J}_{1,0,1})
P^{(2)}_{L_{\pi}}}{\sqrt{1+2 J}\sqrt{L_{\pi}
(1+L_{\pi})}\sqrt{-2+L_{\pi}+{L_{\pi}}^2}}\nonumber\\
&-&\frac{4\sqrt{1+2 L}\sqrt{1+2 L_{\pi}} z
(-1+z^2)D^{L,L_{\pi},J}_{1,-1,1} P^{(3)}_{L_{\pi}}}{\sqrt{1+2
J}\sqrt{L_{\pi}
(1+L_{\pi})}\sqrt{-6+L_{\pi}+{L_{\pi}}^2}\sqrt{-2+L_{\pi}+{{L_{\pi}}^2}}}~,
\nonumber\\
T(L,L_{\pi},J)_{9}&=&\frac{-4\sqrt{1+2 L}\sqrt{1+2 L_{\pi}}
D^{L,L_{\pi},J}_{1,1,-1}P^{(1)}_{L_{\pi}}}{\sqrt{1+2 J}\sqrt{L_{\pi}(1+L_{\pi})}}\nonumber\\
&+&\frac{4\sqrt{1+2 L}\sqrt{1+2
L_{\pi}}(-1+z^2)D^{L,L_{\pi},J}_{1,-1,1} P^{(3)}_{L_{\pi}}}{\sqrt{1+2
J}\sqrt{L_{\pi}(1+L_{\pi})}\sqrt{-6+L_{\pi}+{{L_{\pi}}^2}}\sqrt{-2+L_{\pi}
+{{L_{\pi}}^2}}}~;  \\
L(L,L_{\pi},J)_{1}&=&\frac{2 i \sqrt{2+4 L}\sqrt{1+2 L_{\pi}}
D^{L,L_{\pi},J}_{0,1,1} P_{L_{\pi}}}{\sqrt{1+2 J}}\nonumber\\
&+&\frac{2 i \sqrt{1+2 L}\sqrt{1+2 L_{\pi}} z
(D^{L,L_{\pi},J}_{0,0,1}-D^{L,L_{\pi},J}_{0,1,0})P^{(1)}_{L_{\pi}}}{\sqrt{1+2
J}\sqrt{L_{\pi} (1+L_{\pi})}}~,\nonumber\\
L(L,L_{\pi},J)_{2}&=&\frac{-2 i \sqrt{1+2 L}\sqrt{1+2
L_{\pi}}(D^{L,L_{\pi},J}_{0,0,1}-D^{L,L_{\pi},J}_{0,1,0})P^{(1)}_{L_{\pi}}}{\sqrt{1+2
J}\sqrt{L_{\pi} (1+L_{\pi})}}~,\nonumber\\
L(L,L_{\pi},J)_{3}&=&\frac{-4\sqrt{1+2 L}\sqrt{1+2 L_{\pi}}
(D^{L,L_{\pi},J}_{0,0,1}+D^{L,L_{\pi},J}_{0,1,0})
P^{(1)}_{L_{\pi}}}{\sqrt{1+2 J}\sqrt{L_{\pi} (1+L_{\pi})}}\nonumber\\
&+&\frac{4\sqrt{2+4 L}\sqrt{1+2 L_{\pi}} z D^{L,L_{\pi},J}_{0,1,-1}
P^{(2)}_{L_{\pi}}}{\sqrt{1+2 J}\sqrt{L_{\pi}
(1+L_{\pi})}\sqrt{-2+L_{\pi}+{{L_{\pi}}^2}}}~,\nonumber\\
L(L,L_{\pi},J)_{4}&=&\frac{-4\sqrt{2+4 L}\sqrt{1+2 L_{\pi}}
 D^{L,L_{\pi},J}_{0,1,-1} P^{(2)}_{L_{\pi}}}{\sqrt{1+2 J}\sqrt{L_{\pi} (1+L_{\pi})}\sqrt{-2+L_{\pi}+{{L_{\pi}}^2}}}~,
\end{eqnarray}
where $P^{(n)}_L (z)$ is the $n^{\rm th}$ derivative of the Legendre polynom
$P_L (z)$. The matrices $G_{L_{\pi}}$ and $H_{L_{\pi}}$ can then be expressed 
in terms of the following vectors:
\begin{eqnarray}
G_{L_{\pi}}&=&\left(G_{1},\dots,G_{9}\right) ~, \\
H_{L_{\pi}}&=&\left(H_{1},\dots,H_{4}\right) ~, 
\end{eqnarray}
with
\begin{eqnarray}
G_{1}&=&T(L_{\pi}-1,L_{\pi},L_{\pi}-1)~, \quad
G_{2} = T(L_{\pi}-1,L_{\pi},L_{\pi})~, \quad
G_{3} = T(L_{\pi}+1,L_{\pi},L_{\pi})~,\nonumber\\
G_{4}&=&T(L_{\pi}+1,L_{\pi},L_{\pi}+1)~, \quad
G_{5} = T(L_{\pi}-2,L_{\pi},L_{\pi}-1)~, \quad
G_{6} = T(L_{\pi},L_{\pi},L_{\pi}-1)~,\nonumber\\
G_{7}&=&T(L_{\pi},L_{\pi},L_{\pi})~, \quad
G_{8} = T(L_{\pi},L_{\pi},L_{\pi}+1)~, \quad
G_{9} = T(L_{\pi}+2,L_{\pi},L_{\pi}+1)~, 
\end{eqnarray}
and
\begin{eqnarray}
H_{1}&=&L(L_{\pi}-1,L_{\pi},L_{\pi}-1)~, \quad
H_{2} = L(L_{\pi}-1,L_{\pi},L_{\pi})~, \nonumber\\
H_{3}&=&L(L_{\pi}+1,L_{\pi},L_{\pi})~, \quad
H_{4} = L(L_{\pi}+1,L_{\pi},L_{\pi}+1)~.
\end{eqnarray}

\setcounter{equation}{0}
\section{Two-body to three-body center-of-mass}
\label{app:cmss}

In this appendix we sketch the derivation of the transformation from
the $\gamma$-$d$ center-of-mass (COM) system to the $\gamma$-$N$ COM,
extending the considerations given in \cite{BBLMvK}.
We are interested in the kinematics of the process
$\gamma^* {N_1}{N_2}\rightarrow\pi{N_1}{N_2}$, where the nucleons, $N_1$
and $N_2$, are sewn to the deuteron wavefunctions.  Our 3-body
corrections are evaluated in the $\gamma$-$d$ COM whereas the single
scattering corrections which take into account the scattering of the
photon on the individual nucleons have been calculated in the 
$\gamma$-$N$ COM. It is therefore necessary to construct the Lorentz
transformation which boosts the single-scattering corrections to the
$\gamma$-$d$ COM.

\medskip\noindent
Let $p$ be some four--vector in the  $\gamma$-$d$ COM and $p^*$ the
corresponding four--vector  in the COM of the (second) single
nucleon. These are related by the Lorentz transformation
$p^* = \Lambda (\vec{u}\,) \, p$ with $ \vec{u} = u \, \vec{e}_x$
the velocity. The vector $p$ transforms as
\begin{equation}
\left(\begin{array}{cc}  {p^*_0} \\  {p_\parallel^*} \end{array}\right)=
\left(\begin{array}{ccc}  \gamma & {-\beta\gamma} \\ 
               {-\beta\gamma} & \gamma \end{array}\right)
\left(\begin{array}{ccccc}  {p_0} \\  {p_\parallel} \end{array}\right)~,
\end{equation}
and the transverse directions are of course unaffected. This gives
\begin{equation}
{\vec\beta}=\frac{{{\vec p}_2}+{\vec k}}{k_0 + p_{20}}=
-\frac{{{\vec p}_1}}{k_0 + p_{20}} ~,
\end{equation}
so that $\vec{e}_x = -\hat{p}_1$. The photon energy--momentum 
four--vector thus transforms as 
\beqa
k_0^* &=& \gamma \, \left( k_0 + \beta \vec{k} \cdot \hat{p}_1 \, \right)~,
\nonumber \\
{\vec k}\,^* &=&  \vec{k} -\left( \vec{k} \cdot \hat{p}_1 (1 -\gamma) 
- \gamma \beta {k}_0  \, \right)\, \hat{p}_1~.
\eeqa
Expanded in powers of $1/m$, this reads
\beqa
k_0^* &=&  k_0 + \frac{1}{m}\vec{k} \cdot \vec{p}_1 + {\cal O} (1/m^2)~,
\nonumber \\
{\vec k}\,^* &=&  \vec{k} + \frac{1}{m}  k_0 \, \vec{p}_1 + {\cal O}
(1/m^2)~,
\eeqa
and similarly for the pion energy and three--momentum $(q_0 = \omega,
\vec{q}\,)$. One also has to transform the photon polarization vector.
This is most easily done if one uses the following gauge--invariant
form of the $\gamma^* N \to \pi^0 N$ transition amplitude,
\beq 
{\cal M}_{\rm ss} = \sum_{i=1}^6 \, {\cal O}_{{\rm ss},i} \, F^*_{{\rm
    ss},i}~,
 \eeq
with 
\beqa
{\cal O}_{{\rm ss},1} &=& \vec{\varepsilon}^* \cdot (\hat{k}^* \times
\hat{q}^* )~, \quad 
{\cal O}_{{\rm ss},2}  = \vec{\varepsilon}^* \cdot (\hat{k}^* \times
\hat{q}^* ) \, \vec{S} \cdot (\hat{k}^* \times \hat{q}^* )~, \quad 
{\cal O}_{{\rm ss},3}  = \vec{\varepsilon}^* \cdot (\hat{k}^* \times
(\hat{k}^* \times \vec{S}\,) )~, \quad 
\nonumber \\
{\cal O}_{{\rm ss},4} &=& \vec{\varepsilon}^* \cdot (\hat{k}^* \times
\hat{q}^* ) \, \vec{S} \cdot (\hat{k}^* \times \hat{q}^* )~, \,\, 
{\cal O}_{{\rm ss},5} = \left(  \vec{\varepsilon}^* \cdot \hat{k}^*
- \frac{k^*}{k_0^*} \, \varepsilon_0^* \right) \, \vec{S} \cdot
\hat{k}^*~, \,\,
{\cal O}_{{\rm ss},6} = \left(  \vec{\varepsilon}^* \cdot \hat{k}^*
- \frac{k^*}{k_0^*} \, \varepsilon_0^* \right) \, \vec{S} \cdot
\hat{q}^*~, \nonumber \\
&& 
\eeqa
Since we work in the Coulomb gauge in the $\gamma$-$d$ COM, the
full Lorentz transformation for the polarization vector is given by
\beq
{\varepsilon}_0^* = \gamma \beta \, \vec{\varepsilon} \cdot
\hat{p}_1~, \quad \vec{\varepsilon}\,^* = \vec{\varepsilon} 
-  \vec{\varepsilon} \cdot\hat{p}_1 \,(1-\gamma) \,  \hat{p}_1~.
\eeq 
Expanded in powers of $1/m$, one observes that only the
time--component is modified to leading order,
\beq
{\varepsilon}_0^* = {1\over m} \,  \vec{\varepsilon} \cdot
\vec{p}_1 + {\cal O}(1/m^2) ~, \quad  \vec{\varepsilon}\,^* 
= \vec{\varepsilon}+ {\cal O}(1/m^2) ~.
\eeq

\bigskip
\bigskip
\bigskip


\end{document}